\documentclass[10pt,journal,epsfig]{IEEEtran}
\usepackage{graphicx,makecell}
\usepackage{subfigure,color}
\usepackage{amssymb,setspace}
\usepackage{cite,comment,varwidth}
\usepackage{amsmath,balance}
\usepackage{bm,epstopdf}
\usepackage{algorithm}
\usepackage{algpseudocode}
\makeatletter

\newcommand{\Rmnum}[1]{\expandafter\@slowromancap\romannumeral #1@}
\newcommand{\mv}[1]{\mbox{\boldmath{$ #1 $}}}
\newcommand{\tabincell}[2]{\begin{tabular}{@{}#1@{}}#2\end{tabular}}
\makeatother

\newtheorem{remark}{\underline{Remark}}

\begin{document}
\title{Joint Base Station and IRS Deployment for Enhancing Network Coverage: A Graph-Based Modeling and Optimization Approach}
\author{Weidong Mei, \IEEEmembership{Member, IEEE}, and Rui Zhang, \IEEEmembership{Fellow, IEEE}
\thanks{Part of this work will be presented at the IEEE International Conference on Communications (ICC), Rome, Italy, 2023. {\it (Corresponding author: Rui Zhang)}}
\thanks{W. Mei is with the National Key Laboratory of Science and Technology on Communications, University of Electronic Science and Technology of China, Chengdu 611731, China (e-mail: wmei@uestc.edu.cn).}
\thanks{R. Zhang is with the Chinese University of Hong Kong, Shenzhen, and Shenzhen Research Institute of Big Data, Shenzhen, China 518172 (e-mail: rzhang@cuhk.edu.cn). He is also with the Department of Electrical and Computer Engineering, National University of Singapore, Singapore 117583 (e-mail: elezhang@nus.edu.sg).}}
\maketitle

\begin{abstract}
Intelligent reflecting surface (IRS) can be densely deployed in complex environment to create cascaded line-of-sight (LoS) paths between multiple base stations (BSs) and users via tunable IRS reflections, thereby significantly enhancing the coverage performance of wireless networks. To achieve this goal, it is vital to optimize the deployed locations of BSs and IRSs in the wireless network, which is investigated in this paper. Specifically, we divide the coverage area of the network into multiple non-overlapping cells and decide whether to deploy a BS/IRS in each cell given a total number of BSs/IRSs available. We show that to ensure the network coverage/communication performance, i.e., each cell has a direct/cascaded LoS path with at least one BS, as well as such LoS paths have the average number of IRS reflections less than a given threshold, there is a fundamental trade-off with the deployment cost or the number of BSs/IRSs needed. To optimally characterize this trade-off, we formulate a joint BS and IRS deployment problem based on graph theory, which, however, is difficult to be optimally solved due to the combinatorial optimization involved. To circumvent this difficulty, we first consider a simplified problem with given BS deployment and propose the optimal as well as an efficient suboptimal IRS deployment solution to it, by applying the branch-and-bound method and iteratively removing IRSs from the candidate locations, respectively. Next, an efficient sequential update algorithm is proposed for solving the joint BS and IRS deployment problem. Numerical results are provided to show the efficacy of the proposed design approach and optimization algorithms for the joint BS and IRS deployment. The trade-off between the network coverage performance and the number of deployed BSs/IRSs with different cost ratios is also unveiled.
\end{abstract}
\begin{IEEEkeywords}
	Intelligent reflecting surface (IRS), base station deployment, IRS deployment, network coverage, performance-cost trade-off, graph theory.
\end{IEEEkeywords}

\section{Introduction}
Recent years have witnessed a drastic technological advance in the design, implementation and applications of digitally-controlled metasurfaces, which enable dynamic and flexible manipulation of the impinging electromagnetic waves via tunable passive reflection. In particular, intelligent reflecting surface (IRS) is one such enabling technology that has gained dramatic interests in wireless communications. By adaptively configuring the amplitude and/or phase shifts of a large number of passive reflecting elements, IRS can reshape the wireless channels flexibly to significantly enhance the communication performance in a cost-effective way\cite{wu2020intelligent,di2020smart,basar2019wireless}. Moreover, as compared to the conventional base station (BS) or relay, IRS only reflects the incident signal without the need of any transmit/receive radio frequency (RF) chains, thus considerably reducing the hardware cost and energy consumption, and also making it viable to be densely deployed in the wireless network to improve its communication performance\cite{mei2022intelligent}. Motivated by the great benefits of IRS, its design and enhanced performance have been extensively studied in the literature, from both reflection optimization and channel acquisition perspectives (see e.g., \cite{wu2020intelligent,di2020smart,basar2019wireless,mei2022intelligent,liu2021reconfigurable,pan2022overview,swindlehurst2022channel} and the references therein).

\begin{figure}[!t]
\centering
\includegraphics[width=2.2in]{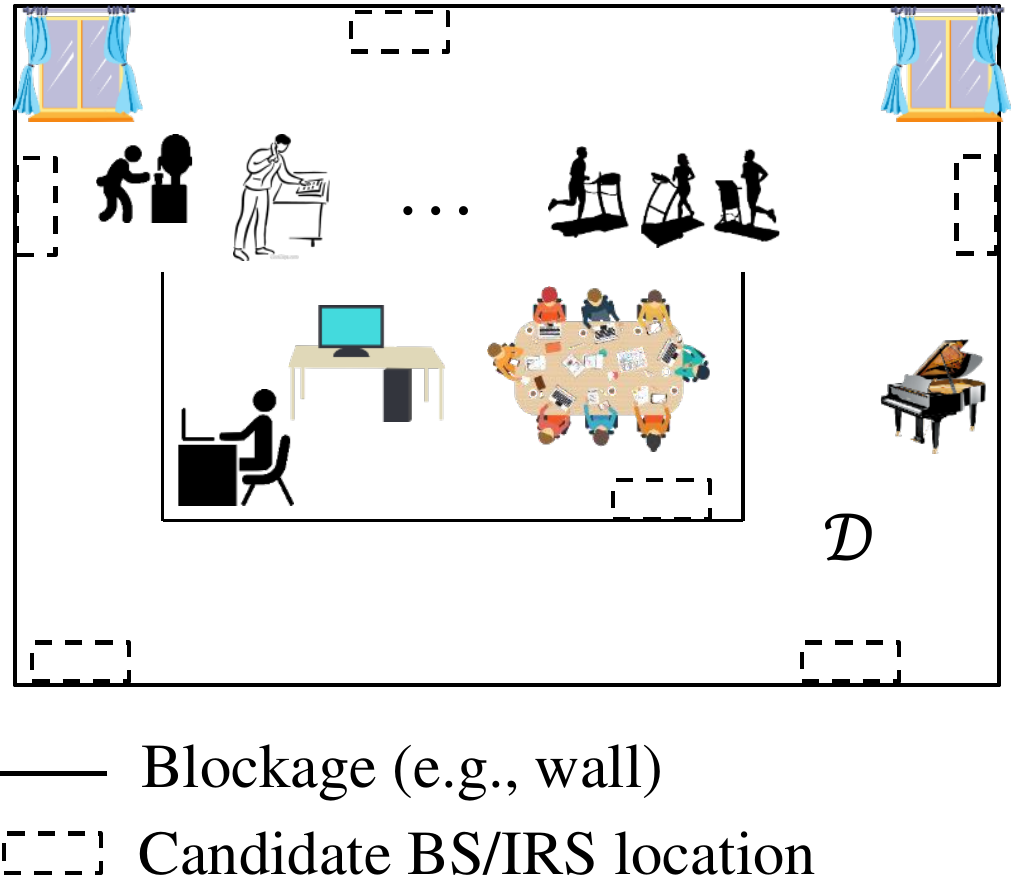}
\DeclareGraphicsExtensions.
\caption{Multi-IRS aided wireless network in a typical indoor environment.}\label{sysmodel}
\vspace{-15pt}
\end{figure}

However, in practice, IRS can only reflect signals from/to its pointing half space and may cause severe signal power loss due to passive reflection. Thus, how to deploy IRSs in wireless networks to optimize the communication performance is a critical problem of high theoretical and practical importance. While most of the existing works on IRS assumed given IRS deployments/locations, some recent works have explored the IRS deployment problem from a communication-theoretic perspective and obtained useful insights. For example, the authors in \cite{zhang2021intelligent} characterized the capacity regions achievable by two IRS deployment strategies with the IRS/IRSs deployed near the BS and each of distributed users, respectively, and showed the superiority of the former over the latter under the same total number of reflecting elements. In \cite{mu2021joint}, a two-timescale design method was proposed, where the IRS deployment and passive reflection were optimized based on the long- and short-term channel knowledge, respectively. In \cite{bai2022robust}, the authors optimized the IRS deployment in a secure wireless communication system to maximize the secrecy rate, in the cases with perfect and partial knowledge on the eavesdropper's location, respectively. In addition, the authors in \cite{lu2021aerial} and \cite{you2021wireless} delved into the deployment problems arising from two new IRS architectures, i.e., aerial IRS and active IRS, respectively. Moreover, the authors in \cite{zeng2021reconfigurable} and \cite{cheng2022ris} further investigated the optimization of the IRS's orientation, in addition to its location. It is shown in \cite{cheng2022ris} that the IRS's rotation may have greater impact on the performance than its location. However, the above works only considered the deployment of one single IRS or multiple co-located IRSs, which may only achieve limited coverage improvement in practical environment with dense obstacles (e.g., the indoor scenario shown in Fig.\,\ref{sysmodel}), where one single IRS or multiple IRSs each reflecting the signal from the BS once only may not be able to establish line-of-sight (LoS) paths between the BS and all desired user locations.

To address this issue, several recent works have investigated the general multi-IRS-reflection aided wireless network. Compared with the conventional single-IRS reflection, multi-IRS reflection provides more available cascaded LoS paths between the BS and user locations, thus further enhancing the IRS-aided coverage performance\cite{mei2022intelligent}. In addition to achieving a higher path diversity, more pronounced cooperative passive beamforming gain can also be reaped from multi-IRS-reflection links, which can compensate the more severe multiplicative path loss\cite{mei2022intelligent}. Motivated by this, the authors in \cite{huang2021multi,zhang2021weighted,liang2022multi} studied the optimal active/passive beamforming design at the BS/IRSs in a given BS-user multi-reflection path. While the authors in \cite{mei2020cooperative,mei2022mbmh,mei2021distributed,mei2022split} studied a more general IRS beam routing problem, which aims to select the optimal multi-IRS-reflection path for each user and jointly optimize the BS/IRS active/passive beamforming in each selected path, such that the end-to-end BS-user channel power gain is maximized. Moreover, the authors in \cite{wang2022novel,tyrovolas2022performance,liu2022intelligent} analyzed several key performance metrics of the general multi-IRS-reflection aided wireless network, such as its ergodic capacity and outage probability, and derived their tractable closed-form performance bounds. Nonetheless, all of the above works\cite{huang2021multi,zhang2021weighted,liang2022multi,mei2020cooperative,mei2022mbmh,mei2021distributed,mei2022split,wang2022novel,tyrovolas2022performance,liu2022intelligent} assumed given BSs' and IRSs' locations; thus, it still remains unknown how to efficiently deploy multiple BSs and IRSs jointly to optimize the wireless network coverage performance in a given region.

To fill this gap, in this paper we investigate a new joint multi-BS and multi-IRS deployment problem for achieving the optimal coverage performance in a wireless network. Our main contributions are summarized as follows.
\begin{itemize}
	\item First, by dividing the network coverage area into multiple non-overlapping cells which contain a set of predetermined candidate locations for deploying BSs/IRSs (see Fig.\,\ref{sysmodel}), we propose a new and general graph-based system model to characterize the network's coverage performance in terms of the total deployment cost of BSs and IRSs. Our analysis reveals that for given BS deployment, there exists a performance-cost trade-off between maximizing the users' communication performance and minimizing the IRS deployment cost.
	\item To further characterize the performance-cost trade-off, we aim to find the optimal locations to jointly deploy BSs and IRSs, such that their total deployment cost is minimized while ensuring the requirement on the network communication performance, i.e., the average number of IRS reflections required for achieving an LoS link between at least one of the BSs and any possible user location in the network needs to be below a target value. However, this problem is a combinatorial optimization problem which is challenging to be solved optimally. To tackle this challenge, we first consider a simplified problem with given locations of the BSs and recast the IRS deployment optimization as a mixed-integer linear programming (MILP) problem, which is then optimally solved via the branch-and-bound (BB) method. To solve this problem more efficiently, we further propose a new successive IRS removal algorithm, which iteratively removes IRSs from the candidate locations while meeting the prescribed communication performance constraint, until convergence is reached.
	\item Next, an efficient sequential update algorithm is proposed for solving the joint BS and IRS deployment problem, where the deployment of multiple BSs is sequentially updated based on the proposed solutions for the simplified problem with given BSs' locations, until no cost reduction is achieved. To ensure the network performance at convergence, we also propose to design the initial BS deployment by maximizing the number of cells that can be directly covered by the BSs without IRSs, which is formulated as another MILP problem and can be optimally solved via the BB method. Simulation results show that our proposed algorithms can yield a flexible performance-cost trade-off and achieve near-optimal performance, thus offering an appealing solution to practical BS and IRS deployment design. It is also revealed that in addition to the performance-cost trade-off, there generally exists an optimal number of BSs deployed to minimize the total BS and IRS deployment cost for any given communication requirement, which depends on the cost ratio between BS and IRS.
\end{itemize}

The rest of this paper is organized as follows. Section \Rmnum{2} presents the system model and the design trade-off in BS/IRS deployment optimization. Section \Rmnum{3} presents the problem formulation for the joint BS and IRS deployment. Sections \Rmnum{4} and \Rmnum{5} present the proposed solutions to this problem in the cases without and with BS deployment optimization, respectively. Section \Rmnum{6} presents the simulation results to show the performance of the proposed solutions and draw useful insights. Finally, Section \Rmnum{7} concludes this paper and discusses future work.
\begin{table*}[htbp]
\centering
\caption{List of Main Symbols}\label{variable}
\resizebox{\textwidth}{!}{
\begin{tabular}{|c|l|c|l|}
\hline
{\bf{Symbol}} & {\bf{Description}} & {\bf{Symbol}} & {\bf{Description}}\\
\hline
$\cal D$ & Region of interest & $N$ & Number of candidate locations or cells \\
\hline
$\cal N$ & Set of cells in $\cal D$ & $\cal B$ & Set of cells deployed with BSs \\
\hline
${\cal I}$ & Set of cells deployed with IRSs & $\mu_{i,j}$ & Binary LoS indicator betweens cells $i$ and $j$ \\
\hline
$G$ & LoS graph for all cells in $\cal D$ & $N_{BS}$ & Number of BSs deployed, $\lvert {\cal B} \rvert = N_{BS}$ \\
\hline
$E$ & Edge set of $G$ & $c({\cal B},{\cal I})$ & Total deployment cost \\
\hline
$\alpha_I$/$\alpha_B$ & Cost per IRS/BS deployed & $W_{i,j}({\cal B},{\cal I})$ & Weight of each edge $(i,j), (i,j) \in E$\\
\hline
$\Gamma_{m,n}$ & \tabincell{l}{Set of all paths from vertex $m$ to vertex $n$\\ in $G$} & $\lambda_{m,n}({\cal B},{\cal I})$ & \tabincell{l}{Minimum IRS number among all LoS paths\\ from cell $m$ to cell $n$}\\
\hline
$\lambda_n({\cal B},{\cal I})$ & \tabincell{l}{Minimum IRS number for cell $n$ from its\\ associated BS} & $\lambda({\cal B},{\cal I})$ & Average minimum IRS number over all cells\\
\hline
$\lambda_0$ & Prescribed maximum value for $\lambda({\cal B},{\cal I})$ & $H$ & Weight of each dummy edge added to $G$\\
\hline
$E_d$ & Set of dummy edges & $\tilde E$ & Edge set of $G$ with dummy edges\\
\hline
$x_{ij}^{mn}$ & \tabincell{l}{Binary indicator of whether edge $(i,j)$ belongs\\ to the selected path from cell $m$ to cell $n$} & $\delta^+_i$/$\delta^-_i$ & \tabincell{l}{Set of outgoing/incoming neighbors of vertex $i$\\ in $G$}\\
\hline
$y_i$ & \tabincell{l}{Binary indicator of whether cell $i$ should be\\ deployed with an IRS} & $\rho_i$ & Auxiliary variable for constraint linearization\\
\hline
$M$ & Large constant for constraint linearization & $w_n$ & Auxiliary variable for constraint linearization\\
\hline
$z_{m,n}$ & Auxiliary variable for constraint linearization & ${\cal B}'$ & BS deployment in the sequential update algorithm\\
\hline
$a_n$ & \tabincell{l}{Binary indicator of whether cell $n$ should be\\ deployed with a BS} & $b_n$ & \tabincell{l}{Binary indicator of whether cell $n$ can be covered\\ by at least one BS}\\
\hline
\end{tabular}}
\end{table*}

The following notations are used in this paper. ${n \choose k} = \frac{n!}{k!(n-k)!}$ denotes the number of ways to choose $k$ elements from a set of $n$ elements. $\lvert A \rvert$ denotes the cardinality of a set $A$. For two sets $A$ and $B$, $A \cap B$ denotes the intersection of $A$ and $B$, $A \cup B$ denotes the union of $A$ and $B$, and $A \backslash B$ denotes the set of elements that belong to $A$ but are not in $B$. $\emptyset$ denotes an empty set. For ease of reference, the main symbols used in this paper are listed in Table \ref{variable}.

\section{System Model and Design Trade-off}
In this section, we first present the system model for the multi-IRS aided wireless network and the involved multi-IRS/BS deployment optimization problem. We also show a fundamental performance-cost  trade-off in this design problem.
\begingroup
\allowdisplaybreaks

\subsection{System Model}
As shown in Fig.\,\ref{sysmodel}, we consider a wireless communication system in a given region ${\cal D}$ with dense obstacles (e.g., a typical indoor scenario with interlaced corridors), which severely block a large portion of communication links over the region, especially for systems operating at higher frequency bands. To enhance the signal coverage, multiple BSs (or access pints (APs)) can be deployed in $\cal D$ to establish LoS links with as many user locations as possible in $\cal D$. However, due to the dense and scattered obstacles, this may require deploying an excessively large number of BSs to achieve the global LoS coverage over ${\cal D}$, thus resulting in practically unaffordable deployment cost and power consumption, as well as the increasing difficulty in managing their potential interference. To enhance the network coverage in a more efficient manner, multiple passive IRSs can be properly deployed in ${\cal D}$ to create virtual LoS paths between any two locations in $\cal D$, thereby significantly reducing the number of BSs required. For simplicity, we assume that a number of candidate locations, denoted by $N$, have been identified in $\cal D$, each of which can be deployed with either one BS or one IRS, where the IRSs can help create LoS links with neighboring BSs/IRSs, and/or certain user locations in $\cal D$. Moreover, for the purpose of exposition, we assume that the number of deployed BSs is fixed, denoted as $N_{BS}$.\footnote{Our results apply to any value of $N_{BS}$, while its effects will be shown later in Section \ref{sim3} via simulation.} In practice, such candidate locations for BS/IRS deployment can be determined via e.g., the ray-tracing approach based on the known topology of the region\cite{fuschini2015ray}.

To facilitate the joint BS and IRS deployment design, we discretize the region ${\cal D}$ into multiple non-overlapping cells, each containing at most one candidate BS/IRS deployment location.\footnote{To achieve this purpose, we can first divide the region into multiple small-size grids and merge several adjacent grids into a larger-size cell if any two grids inside it can achieve LoS propagation with each other. The LoS availability can be determined in practice based on ray tracing.} It is also assumed that LoS paths can be achieved between the candidate deployment location (if any) and any other possible user locations in each cell. As such, it is obvious that if each cell has a candidate deployment location and BSs are deployed at all these candidate locations, then the global network coverage can be achieved, as we can always find a direct LoS path (without the need of any IRS reflection) between any possible user location\footnote{For ease of exposition, we consider a 2D coverage problem in this paper, i.e., the user locations are at a fixed height off the ground, while the results of this paper can be extended to solve the general 3D coverage problem.} in $\cal D$ and the BS deployed in its located cell. However, due to the high BS deployment cost, it is usually impractical to deploy a BS at each of all these candidate locations; as such, only a subset of locations are chosen for deploying BSs to reduce the total cost, i.e., $N_{BS} \ll N$, by leveraging the LoS links among adjacent cells and the multi-IRS signal reflections (to be shown later). Similarly, for any given BS deployment, it may suffice to only deploy IRSs in a subset of remaining candidate locations to achieve a global LoS coverage in the network.

\begin{figure}[!t]
\centering
\includegraphics[width=3.5in]{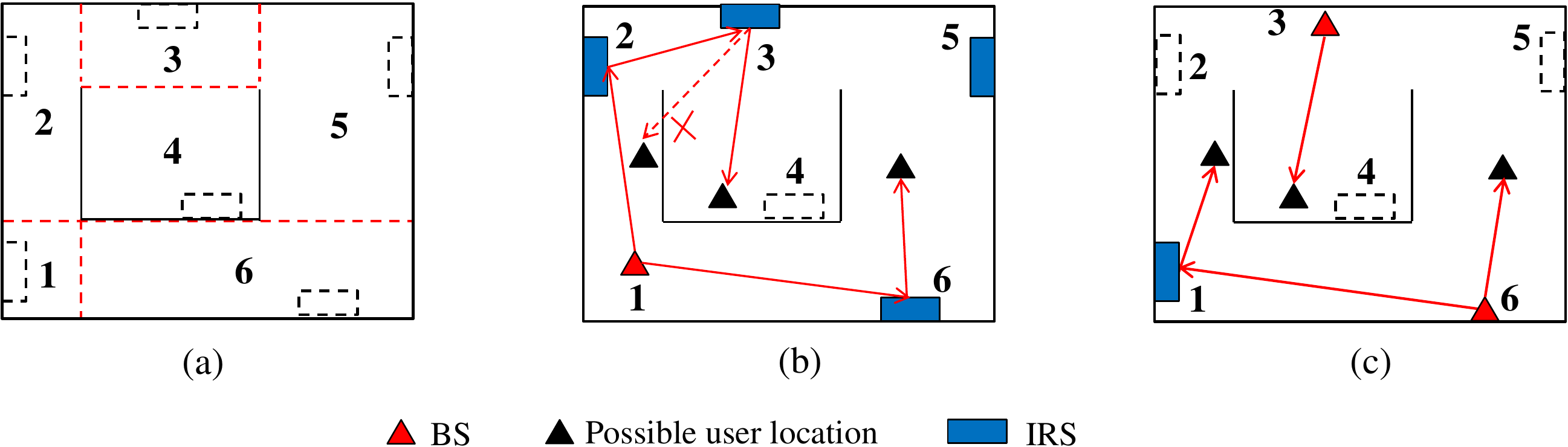}
\DeclareGraphicsExtensions.
\vspace{-6pt}
\caption{Illustrations for (a) the discretization of the region in Fig.\,\ref{sysmodel}, (b) LoS paths created by IRSs deployed at partial candidate locations with ${\cal B}=\{1\}$ and (c) those with ${\cal B}=\{3,6\}$.}\label{graph}
\vspace{-12pt}
\end{figure}
Let the cells that partition the region $\cal D$ be denoted by the set ${\cal N}=\{1,2,\cdots,N\}$, where for ensuring that a global network coverage is always feasible, we have assumed that the number of cells equals that of BS/IRS deployment locations, $N$, i.e., each cell contains one BS/IRS candidate deployment location, as shown in Fig.\,\ref{graph}(a); while the results of this paper can be easily extended to the case where some cells do not contain any candidate deployment locations. Accordingly, let ${\cal B} \subseteq \cal N$ and ${\cal I} \subseteq \cal N$ denote the sets of cells deployed with BSs and IRSs, respectively, with $\lvert {\cal B} \rvert=N_{BS}$ and ${\cal B} \cap {\cal I} = \emptyset$. Moreover, we define a set of binary variables $\mu_{i,j}, i \ne j, i,j \in {\cal N}$ to indicate the {\it direct} LoS path availability between any two different cells $i$ and $j$ (we assume $\mu_{i,i}=1, \forall i$ without loss of generality). In particular, $\mu_{i,j}=1$ holds if and only if (iff) the candidate BS/IRS location in cell $i$ can achieve a direct LoS path with any possible user locations as well as the candidate BS/IRS location in cell $j$. Note that $\mu_{i,j}=1$ may not necessarily lead to $\mu_{j,i}=1$, as this depends on the candidate BS/IRS location in cell $j$ (e.g., $\mu_{2,3}=1$ but $\mu_{3,2}=0$ as shown in Fig.\,\ref{graph}(b)). Moreover, the user/IRS in any cell $j$ may be covered by a nearby or remote BS in cell $i, i \ne j$ via a direct LoS link (without the need of any additional IRS's reflection) and a cascaded LoS link formed by a number of other IRSs deployed in cells $k$ with $k \ne i$ and $k \ne j$, respectively. For example, in Fig.\,\ref{graph}(b) with a single BS deployed in cell 1, we have $\mu_{1,4}=0$ but $\mu_{1,2}=\mu_{2,3}=\mu_{3,4}=1$. Hence, any user in cell 4 can be served by the BS over a multi-IRS LoS reflection link via the two IRSs deployed in cells 2 and 3, respectively\footnote{Note that the user in cell 4 may also be served by the IRS deployed in this cell (if any) for improving its communication performance. However, this is irrelevant to our considered IRS deployment design in this paper, which mainly focuses on the network coverage performance.}. On the other hand, in Fig.\,\ref{graph}(c) with two BSs respectively deployed in cells 3 and 6, any user in cell 4 now can be served by the BS in cell 3 over a direct LoS link. In this paper, we focus on the downlink coverage problem, while the results can be similarly applied to the uplink case as well, since for any given user location, a direct/multi-IRS-reflected LoS path from each BS to it in the downlink is also available for its uplink communication to this BS in practice. It is also assumed that the LoS indicators $\mu_{i,j}$'s are known {\it a priori} for the region $\cal D$ of interest via e.g., ray-tracing.

Next, we model the joint BS and IRS deployment problem based on a directed LoS graph $G=(V,E)$ for all cells in $\cal D$, where the vertex set $V$ contains all the $N$ cells, i.e., $V={\cal N}$, and the edge set is given by $E=\{(i,j)|\mu_{i,j}=1, i \ne j\}$, i.e., there is an edge from one vertex $i$ to another vertex $j$ iff $\mu_{i,j}=1$. By this means, we can establish a one-to-one mapping between any direct/cascaded LoS path in $\cal D$ and a path in $G$. Fig.\,\ref{example}(a) depicts one graph $G$ generated based on the cells in Fig.\,\ref{graph}(a) and their pairwise binary LoS indicators $\mu_{i,j}$'s. For example, the cascaded LoS path in Fig.\,\ref{graph}(b) from the BS to cell 4 via the IRSs deployed in cells 2 and 3 corresponds to the path from vertex 1 to vertex 4 via vertices 2 and 3 in $G$. Hence, in the sequel of this paper, we interchangeably use vertices and cells and a path can refer to both an LoS path in $\cal D$ and its corresponding path in $G$ without ambiguity.
\begin{figure}[!t]
\centering
\includegraphics[width=3.5in]{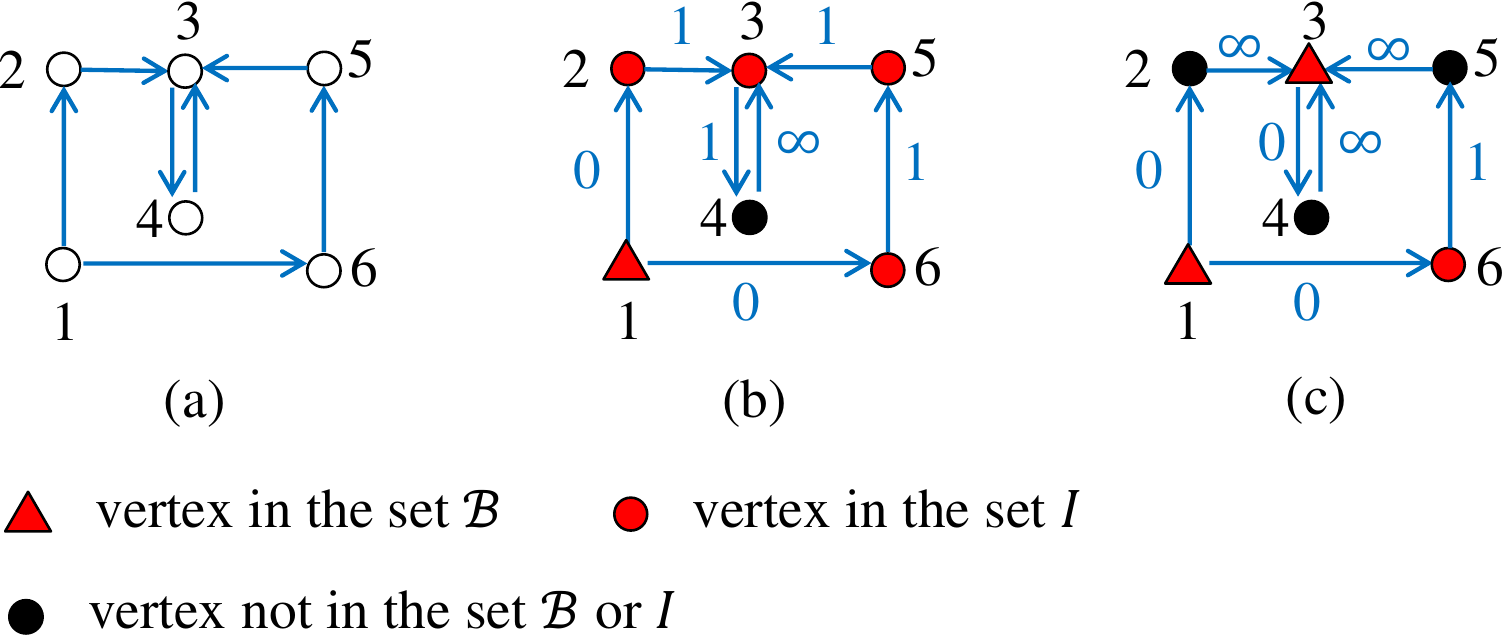}
\DeclareGraphicsExtensions.
\vspace{-12pt}
\caption{Illustrative examples of (a) graph $G$ and its weight assignments under two different BS/IRS deployments shown in (b) and (c), respectively.}\label{example}
\vspace{-12pt}
\end{figure}

\subsection{Design Trade-off}\label{pm}
Next, we show a fundamental trade-off between the cost and coverage performance in our considered joint BS and IRS deployment problem. In particular, we consider that the total BS/IRS deployment cost is proportional to the number of BSs/IRSs deployed, i.e., $N_{BS}$ and $\lvert {\cal I} \rvert$. As such, the total deployment cost is given by
\begin{equation}\label{cost}
c({\cal B},{\cal I})=\alpha_B N_{BS} + \alpha_I\lvert {\cal I} \rvert,
\end{equation}
where $\alpha_B$/$\alpha_I$ denotes the cost per BS/IRS deployed. It is worth noting that in practice, it usually holds that $\alpha_B \gg \alpha_I$ due to the much lower hardware cost and energy consumption of passive IRSs as compared to active BSs/APs. Since $N_{BS}$ is assumed to be given, $c({\cal B},{\cal I})$ only depends on the number of IRSs deployed, i.e., $\lvert {\cal I} \rvert$.

For any given BS and IRS deployment ${\cal B}$ and ${\cal I}$, the coverage performance in each cell $i, i \in {\cal B}^c \triangleq {\cal N} \backslash {\cal B}$, depends on whether an LoS path can be established from at least one of the BSs in $\cal B$ to it given the IRSs deployed in ${\cal I}$. Moreover, if such an LoS path exists, the communication performance of any user in cell $i$ also depends on the number of IRSs involved (if any) for creating this cascaded LoS path with its associated BS, which is desired to be as small as possible to avoid higher multiplicative path loss due to multi-IRS reflections\cite{mei2022intelligent,mei2020cooperative,mei2022mbmh}. Thus, we consider the number of (reflecting) IRSs in each LoS path from a BS in $\cal B$ to cell $i, i \in {\cal B}^c$, as the communication performance metric of users in cell $i$ via this LoS path. It is worth noting that inter-user interference may exist in the considered system due to the BSs' transmission and the IRSs' uncontrollable scattering, which can be mitigated via proper resource allocation and BS/IRS beamforming, as well as sufficient separation among the LoS paths associated with different users\cite{mei2022mbmh}. As such, it is not accounted for in our BS/IRS deployment design that is performed prior to the actual communications.

Then, to help count the number of IRSs involved in any IRS-aided LoS reflection path from each BS to a cell with given ${\cal I}$, we define the following weight function for each edge $(i,j)$ in $G$ in terms of ${\cal B}$ and ${\cal I}$, i.e.,
\begin{equation}\label{wt}
W_{i,j}({\cal B},{\cal I}) = \begin{cases}
	0 &{\text{if}}\;\;i \in {\cal B}, j \in {\cal B}^c\\
	1 &{\text{if}}\;\;i \in {\cal I}, j \in {\cal B}^c\\
	\infty &{\text{otherwise}},
	\end{cases}
\qquad(i,j) \in E.
\end{equation}
In particular, since any LoS path should start from a BS in $\cal B$ without the need of any IRS, we set the weight of each edge starting from a vertex in $\cal B$ to zero, as shown in the first case of (\ref{wt}). While if an LoS path goes through an intermediate vertex $i, i \in \cal I$ (IRS) to another vertex $j, j \in {\cal B}^c$ (IRS/user), this indicates that the signal from the BS in this path will be reflected by the IRS deployed in cell $i$. Thus, we set the weight of each edge starting from vertex $i, i \in \cal I$ to one, as shown in the second case of (\ref{wt}). Finally, if vertex $j$ corresponds to a cell deployed with a BS, i.e., $j \in \cal B$, then the edge $(i,j)$ cannot be used to form any desired LoS path, since the cells deployed with BSs can neither reflect the signal as an intermediate node nor receive it as a destination node (as the downlink scenario is considered). In addition, if neither the BS nor an IRS is deployed in cell $i$, then cell $i$ should not be selected as an intermediate cell by any desired LoS path. As such, in both of the above cases, the edge $(i,j)$ should be removed under the BS/IRS deployment ${\cal B}$ and $\cal I$. However, to keep the topology of $G$ and ease the IRS number counting, we set their weights to infinity equivalently, as shown in the third case of (\ref{wt}). For illustration, Figs.\,\ref{example}(b) and \ref{example}(c) depict the weight assignments for the graph $G$ in Fig.\,\ref{example}(a) under the BS/IRS deployment in Figs.\,\ref{graph}(b) and \ref{graph}(c), respectively.

For any given BS deployment ${\cal B}$ and IRS deployment $\cal I$, let $\Gamma_{m,n}$ denote the set of all LoS paths from vertex $m, m \in \cal B$ to vertex $n, n \in {\cal B}^c$ in $G$ and $\Omega=\{m,a_1,a_2,\cdots,a_L,n\} \in \Gamma_{m,n}$ denote one of the paths in $\Gamma_{m,n}$ if $\Gamma_{m,n} \ne \emptyset$, where $L$ denotes the number of its intermediate vertices. Assuming $a_0=m$ and $a_{L+1}=n$ for convenience, the sum of weights of the path $\Omega$ is given by
\begin{equation}
\lambda(\Omega,{\cal B},{\cal I})=\sum\limits_{l=0}^L W_{a_l,a_{l+1}}({\cal B},{\cal I}).
\end{equation}
Evidently, if $\lambda(\Omega,{\cal B},{\cal I})=\infty$, then at least one of the weights $W_{a_l,a_{l+1}}({\cal B},{\cal I})$'s is equal to infinity. As such, the BS in cell $m$ cannot achieve the LoS coverage in cell $n$ via the path $\Omega$ under the IRS deployment $\cal I$. While if $\lambda(\Omega,{\cal B},{\cal I})<\infty$, it must hold that $a_l \in {\cal I}, 1 \le l \le L$ and thus the LoS coverage can be achieved via the path $\Omega$. In this case, it follows from (\ref{wt}) that $\lambda(\Omega,{\cal B},{\cal I})=L$, i.e., $\lambda(\Omega,{\cal B},{\cal I})$ is equal to the number of IRSs involved in the path $\Omega$, which is $L$.

Based on the above, the minimum (reflecting) IRS number among all LoS paths from the BS in cell $m$ to cell $n$ is obtained as
\begin{equation}\label{min1}
\lambda_{m,n}({\cal B},{\cal I})=\min\limits_{\Omega \in \Gamma_{m,n}}\lambda(\Omega,{\cal B},{\cal I}), n \in {\cal B}^c, m \in {\cal B},
\end{equation}
which can be efficiently calculated by invoking the classical shortest path algorithm in graph theory (e.g., Dijkstra algorithm\cite{west1996introduction}) on $G$ with the weights given in (\ref{wt}). On the other hand, if $\Gamma_{m,n}=\emptyset$, i.e., there is no path from vertex $m, m \in \cal B$ to vertex $n, n \in {\cal B}^c$ in $G$ (e.g., from vertex 3 to vertex 2 or 5 in Fig.\,\ref{example}(c)), we should set $\lambda_{m,n}({\cal B},{\cal I})=\infty$. Then, among all BSs in $\cal B$, any user in cell $n$ should be associated with the one having the minimum IRS number to it, so as to minimize the multiplicative path loss\footnote{In practice, the IRS in one cell may be involved in the cascaded LoS paths for multiple BSs/cells. In this case, we can divide the IRS into multiple subsurfaces to reflect the incident signals to different nearby IRSs/cells in different paths, or schedule the reflection of the IRS with them over orthogonal time slots.}. The corresponding minimum IRS number for cell $n$ is thus given by
\begin{equation}\label{min2}
\lambda_n({\cal B},{\cal I})=\min\limits_{m \in {\cal B}}\lambda_{m,n}({\cal B},{\cal I}), n \in {\cal B}^c.
\end{equation}
For convenience, we further set $\lambda_n({\cal B},{\cal I})=0$ if $n \in {\cal B}$, as any user in cell $n, n \in \cal B$ can be directly covered by the BS deployed in this cell without the need of IRS. It is noted that if $\lambda_n({\cal B},{\cal I}) < \infty, \forall n \in {\cal N}$, then each user in cell $n$ can achieve an LoS link (direct or single-/multi-IRS reflection) with at least one BS in $\cal B$.\footnote{Note that if $\lambda_n({\cal B},{\cal I}) \le 1, \forall n \in {\cal N}$, then the global LoS coverage can be achieved with only direct and single-IRS-reflection LoS links. However, this generally requires deploying a large number of BSs, especially in a complex environment, which may incur an excessively high deployment cost.} As a result, a global LoS coverage can be achieved in the wireless network with the BS deployment ${\cal B}$ and IRS deployment $\cal I$. In this case, we define the average minimum IRS number over all cells in ${\cal N}$ as
\begin{equation}\label{sum}
\lambda({\cal B},{\cal I})=\frac{1}{N}\sum\limits_{n \in \cal N}\lambda_n({\cal B},{\cal I}),
\end{equation}
which is desired to be smaller for achieving better communication performance of the users in the region $\cal D$. Hence, for any given ${\cal B}$ and $\cal I$, the performance metric in (\ref{sum}) can capture both the feasibility of the global LoS coverage (by checking whether $\lambda({\cal B},{\cal I})<\infty$ is met) and the users' communication performance (at the network level).

\begin{remark}
It should be mentioned that the performance metric in (\ref{sum}) only aims to capture the users' communication performance at the network level without the actual channel or user location information available at the stage of BS/IRS deployment. The actual communication performance can be further optimized after their deployment based on the users' channel knowledge or real-time beam training and selection at the BS/IRS. Nonetheless, the proposed graph-based modeling approach can be extended to consider other finer performance metrics, by accounting for e.g., the large-scale channel condition among the BS and IRSs, in the weight functions of the LoS graph in (\ref{wt}).
\end{remark}

It is worth noting that for any given BS deployment $\cal B$, $\lambda({\cal B},{\cal I})$ can be minimized by deploying IRSs in all other cells in ${\cal N}$, i.e., ${\cal I}={\cal B}^c$ and $\lambda({\cal B},{\cal I}) \ge \lambda({\cal B},{\cal B}^c), \forall {\cal I} \subseteq {\cal B}^c$, as this results in the largest possible size of each $\Gamma_{m,n}$ in (\ref{min1}); however, the maximum IRS deployment cost in (\ref{cost}) occurs as well. In general, with increasing/decreasing the number of BSs/IRSs deployed (or the deployment cost $c({\cal B},{\cal I})$), $\lambda({\cal B},{\cal I})$ may decrease/increase due to the enhanced/reduced LoS path diversity from any BS to any cell. Thus, there exists a fundamental trade-off between minimizing $c({\cal B},{\cal I})$ and minimizing $\lambda({\cal B},{\cal I})$ in jointly optimizing ${\cal B}$ and $\cal I$ for our considered joint BS and IRS deployment problem.

\section{Problem Formulation}\label{pf}
To characterize the optimal performance-cost trade-off, in this section, we aim to optimize the BS and IRS deployment (i.e., their numbers and deployed location) $\cal B$ and $\cal I$, to minimize the total deployment cost $c({\cal B},{\cal I})$, subject to the constraint on the network coverage performance, in terms of the average minimum IRS number $\lambda({\cal B},{\cal I})$ given in (\ref{sum}). The associated optimization problem is formulated as
\begin{align}
{\text{(P1)}}\; \mathop {\min}\limits_{{\cal B},{\cal I}}\;\;&\lvert {\cal I} \rvert \nonumber\\
\text{s.t.}\;\;& {\cal B} \subseteq {\cal N}, {\cal I} \subseteq {\cal N}, \label{op1a}\\
&\lvert {\cal B} \rvert = N_{BS}, \;{\cal B} \cap {\cal I} = \emptyset, \label{op1b}\\
&\lambda({\cal B},{\cal I}) \le \lambda_0,\label{op1c}
\end{align}
where the constant $\alpha_B N_{BS}$ and scalar $\alpha_I$ are omitted in the objective function, and $\lambda_0 \ge 0$ denotes a prescribed maximum value for the average minimum IRS number. By varying the value of $\lambda_0$, we can obtain the performance-cost trade-off region between $\lvert {\cal I} \rvert$ and $\lambda({\cal B},{\cal I})$, as will be shown in Section \ref{sim}. Note that there should exist a minimum $\lambda_0$ to ensure that (P1) is feasible in general. However, if the number of BSs, $N_B$ is sufficiently large, each user may be served by at least one BS over a direct LoS link without the need of IRS reflections. In this case, we have ${\cal I}=\emptyset$ and $\lambda({\cal B},\emptyset)=0$, thus making (P1) always feasible regardless of $\lambda_0$.

It is worth mentioning that the joint BS and IRS deployment problem (P1) is different from existing BS deployment problems for enhancing the network coverage (see e.g., \cite{liu2012femtocell,gonzalez2011base,bi2015placement}), due to the new coupling of BS and IRS deployment. Moreover, (P1) is also different from the conventional relay deployment problems in wireless networks, as studied in e.g., \cite{lloyd2006relay,misra2009constrained,han2009fault}. This is because the latter problem mainly aims to deploy relays to serve users at some given locations; while our problem focuses on enhancing the entire network coverage in a given region, which needs to cater to all possible user locations therein.

However, (P1) is a non-convex combinatorial optimization problem, which is challenging to be optimally solved in general. One straightforward approach to optimally solve (P1) is by enumerating all possible BS and IRS deployments, i.e., first choosing $N_{BS}$ cells (out of $N$ cells) to deploy BSs and then determining whether or not to deploy an IRS in each of the remaining $N-N_{BS}$ cells in ${\cal B}^c$. However, this incurs an exorbitant complexity of ${N \choose N_{BS}} 2^{N-N_{BS}}$ (which is in the order of $10^{10}$ for our considered numerical example with the number of cells, $N=25$, in Section \ref{sim} and $N_{BS}=3$), which exponentially increases with $N$ and thus is not applicable to a large region with large $N$ values required in practice. To cope with this challenging problem, we first consider a simplified version of (P1) by fixing the locations of BSs, and propose optimal and efficient suboptimal algorithms to solve it in Section \Rmnum{4}, which will also be used to solve the general problem (P1) in Section \Rmnum{5}.

\section{Proposed Solution to (P1) with Given BS Locations}\label{sol1}
In this section, we consider the case that the locations of deployed BSs (as specified by $\cal B$) are fixed and given. In this case, (P1) can be simplified as
\begin{align}
{\text{(P2)}}\; \mathop {\min}\limits_{{\cal I}}\;\;& \lvert {\cal I} \rvert \nonumber\\
\text{s.t.}\;\;& {\cal I} \subseteq {\cal B}^c, \;\lambda({\cal B},{\cal I}) \le \lambda_0.\label{op2}
\end{align}
For convenience, we assume in this section that for any cell $i, i \in {\cal B}^c$, it can achieve an LoS path (direct or IRS-reflected) with at least one BS in $\cal B$ if IRSs are deployed in all cells in ${\cal B}^c$, i.e., ${\cal I}={\cal B}^c$. As a result, we have $\lambda({\cal B},{\cal B}^c) < \infty$ and consider $\lambda_0 \ge \lambda({\cal B},{\cal B}^c)$ to ensure the feasibility of (P2). Nonetheless, if the number of BSs deployed, $N_{BS}$, is insufficient, there generally exists a minimum $\lvert {\cal I} \rvert$ below which $\lambda({\cal B},{\cal I}) < \infty$ cannot hold. In the following, we show that (P2) can be optimally solved via the BB algorithm by recasting it as an MILP problem. To reduce the computational complexity, a suboptimal successive IRS removal algorithm of polynomial complexity is also proposed to solve (P2) more efficiently.

\subsection{Optimal Solution by the BB Algorithm}
Our proposed optimal algorithm to solve (P2) includes the following three main operations.

{\it 1) Add dummy edges:} First, to facilitate our problem reformulation of (P2), we add a ``dummy'' edge with an infinitely large weight $H, H \rightarrow \infty$, between any two BSs $i$ and $j$ in $\cal B$, $i \ne j$. The set of dummy edges is thus given by
\begin{equation}\label{dummy}
E_d = \{(i,j)| i,j \in {\cal B}, i \ne j\}.
\end{equation}
By this means, it can be ensured that there exists at least one path from any vertex $m, m \in \cal B$ to any vertex $n, n \in {\cal B}^c$ in $G$. For example, in Fig.\,\ref{example}(c), vertex 3 can now reach vertex 2 or 5 by first going to vertex 6 through the dummy edge between them. Nonetheless, as $H \rightarrow \infty$, any path going through a dummy edge will never be selected to serve users, as this will violate the constraint in (\ref{op2}) on the average minimum IRS number. Thus, the optimal solution to (P2) is unaffected after adding the dummy edges. With the added dummy edges, the edge set of $G$ becomes $\tilde E=E \cup E_d$.


{\it 2) Reformulate (P2) as an integer programming:} Next, we define a set of binary variables ${\mv X}=\{x_{ij}^{mn}\}, m \in {\cal B}, n \in {\cal B}^c, (i,j) \in \tilde E$ to specify a path from any vertex $m, m \in \cal B$ to any vertex $n, n \in {\cal B}^c$. In particular, we set $x_{ij}^{mn}=1$ if the edge $(i,j)$ belongs to this path. Otherwise, we set $x_{ij}^{mn}=0$. Then, it can be shown that the following {\it flow conservation} constraints can specify any valid path from vertex $m$ to vertex $n$ in $G$,
\begin{equation}\label{flow}
\sum\limits_{j \in \delta^+_i}{x_{ij}^{mn}}-\sum\limits_{j \in \delta^-_i}{x_{ji}^{mn}}=
\begin{cases}
1, &{\text{if}}\;\; i=m\\
-1, &{\text{if}}\;\; i=n\\
0, &{\text{otherwise}},
\end{cases}
\end{equation}
where $\delta^+_i=\{j|(i,j) \in \tilde E\}$ and $\delta^-_i=\{j|(j,i) \in \tilde E\}$ denote the set of outgoing and incoming neighbors of vertex $i$, respectively. For example, in Fig.\,\ref{example}(b), consider the path from the BS in cell 1 to cell 4 (i.e., $m=1$ and $n=4$) via cells 2 and 3. Thus, we have $x_{12}^{14}=1$, $x_{23}^{14}=1$, and $x_{34}^{14}=1$, while the binary variables associated with other edges in $G$ are all equal to zero. It can be verified that all the conditions in (\ref{flow}) can be satisfied for this path.

As such, for any path specified by (\ref{flow}), if there is no dummy edge involved, the number of reflecting IRSs in this path should be equal to the number of its constituent edges subtracted by one (corresponding to the edge starting from the BS in cell $m$), i.e., $\lambda_{m,n}({\mv X})=\sum\nolimits_{(i,j) \in {\tilde E}} x_{ij}^{mn}-1$. For example, the number of IRSs in the above path in Fig.\,\ref{example}(b) is equal to $\lambda_{1,4}({\mv X})=x_{12}^{14}+x_{23}^{14}+x_{34}^{14}-1=2$. Taking into account the effect of dummy edges, we define the effective number of reflecting IRSs in any path as
\begin{equation}\label{wtsum}
\lambda_{m,n}({\mv X})=\sum\limits_{(i,j) \in {\tilde E}} c_{ij}x_{ij}^{mn}-1,
\end{equation}
where
\[c_{ij}=\begin{cases}
H, &{\text{if}}\; (i,j) \in E_d\\
1, &{\text{if}}\; (i,j) \in E,
\end{cases}\]
and the average minimum IRS number over all cells in ${\cal N}$ can be expressed as
\begin{align}
\lambda({\mv X})=\frac{1}{N}\sum\limits_{n \in \cal N}\min\limits_{m \in {\cal B}}\lambda_{m,n}({\mv X}).
\end{align}

Next, we express the number of IRSs deployed with the binary variables in $\mv X$. Note that an IRS should be deployed in a cell $i, i \in {\cal B}^c$ if this cell is involved in at least one of the specified paths from all BSs in $\cal B$ to all other cells in ${\cal B}^c$. This is equivalent to the case that at least one of $x_{ij}^{mn}$'s over $m \in {\cal B}$, $n \in {\cal B}^c$ and $j \in \delta^+_i$ is equal to one, i.e.,
\begin{equation}
\sum\limits_{m \in {\cal B}}\sum\limits_{n \in {\cal B}^c}\sum\limits_{j \in \delta^+_i} x_{ij}^{mn} \ge 1.
\end{equation}
Otherwise, there is no need to deploy an IRS in cell $i$. Accordingly, we define
\begin{equation}\label{yi}
y_i = \min \biggl\{1,\sum\limits_{m \in {\cal B}}\sum\limits_{n \in {\cal B}^c}\sum\limits_{j \in \delta^+_i} x_{ij}^{mn}\biggl\}, i \in {\cal B}^c.
\end{equation}
Then, it is easy to see that cell $i$ should be deployed with an IRS iff $y_i=1$. Hence, the total number of IRSs deployed is given by $\lvert {\cal I} \rvert = \sum\nolimits_{i \in {\cal B}^c} y_i$.

Let ${\mv Y}$ denote the ensemble of $y_i, i \in {\cal B}^c$. Based on the above, (P2) can be reformulated as
\begin{align}
{\text{(P2.1)}}\; &\mathop {\min}\limits_{{\mv X}, {\mv Y}}\;\; \sum\limits_{i \in {\cal B}^c} y_i \nonumber\\
\text{s.t.}\;\;& \min \biggl\{1,\sum\limits_{m \in {\cal B}}\sum\limits_{n \in {\cal B}^c}\sum\limits_{j \in \delta^+_i} x_{ij}^{mn}\biggl\} \le y_i, \forall i \in {\cal B}^c,\label{op2-1a}\\
& \frac{1}{N}\sum\limits_{n \in \cal N}\min\limits_{m \in {\cal B}}\lambda_{m,n}({\mv X}) \le \lambda_0, \label{op2-1b}\\
&{\text{(\ref{flow})}}, x_{ij}^{mn} \in \{0,1\}, \forall (i,j) \in {\tilde E}, m \in {\cal B}, n \in {\cal B}^c.
\end{align}
Note that the equalities in (\ref{yi}) have been rewritten as the inequalities in (\ref{op2-1a}). This is because (\ref{op2-1a}) must hold with equality at the optimality of (P2.1); otherwise, we can decrease each $y_i$ until the equality holds, while achieving a smaller objective value of (P2.1). However, (P2.1) turns out to be a non-convex mixed-integer nonlinear programming (MINLP) problem, which is NP-hard and difficult to be optimally solved. In fact, even by relaxing all integer variables in ${\mv X}$ into their continuous counterparts in (P2.1), i.e., $0 \le x_{ij}^{mn} \le 1, \forall (i,j),m,n$, (P2.1) is still a non-convex optimization problem owing to constraints (\ref{op2-1a}) and (\ref{op2-1b}), since their left-hand sides (LHSs) involve the pointwise minimum of a set of affine functions, thus being a concave (instead of convex) function in $\mv X$.
\begin{remark}
It should be mentioned that in (P2.1), we do not restrict that the selected path specified by $x_{ij}^{mn}$'s in (\ref{flow}) is the minimum-weight path among all paths from vertex $m$ to vertex $n$ in $G$, which is different from (\ref{min1}). However, we argue that this relaxation is without loss of optimality, and the optimal value of (P2.1) is identical to that of (P2). This is because for any IRS deployment, if the selected path can satisfy the constraint on the minimum IRS number in (\ref{op2-1b}), the minimum-weight path (if it is different from the selected one) must also satisfy it.
\end{remark}

{\it 3) Linearize (P2.1) as an MILP:} To deal with problem (P2.1), we linearize the two constraints (\ref{op2-1a}) and (\ref{op2-1b}) by invoking the ``big-$M$'' method\cite{belotti2013mixed} in solving MINLP. Specifically, for constraint (\ref{op2-1a}), we introduce a set of binary variables $\rho_i \in \{0,1\}, i \in {\cal B}^c$ and define $M$ as a sufficiently large constant with $M \rightarrow \infty$. Then, constraint (\ref{op2-1a}) can be equivalently recast as the following linear constraints,
\begin{align}
&\rho_i \in \{0,1\}, \forall i \in {\cal B}^c, \label{eq1}\\
& 1-M(1-\rho_i) \le y_i, \forall i \in {\cal B}^c,\label{eq2}\\
& \sum\limits_{m \in {\cal B}}\sum\limits_{n \in {\cal B}^c}\sum\limits_{j \in \delta^+_i} x_{ij}^{mn}-M\rho_i \le y_i, \forall i \in {\cal B}^c. \label{eq3}
\end{align}
It is noted that the binary variable $\rho_i$ is introduced to control the active states of (\ref{eq2}) and (\ref{eq3}). In particular, if $\rho_i=1$ or $\rho_i=0$, only constraint (\ref{eq2})/(\ref{eq3}) would take effect, while the other becomes inactive. As a result, to minimize the sum of $y_i$'s, each optimal $\rho_i$ should choose between 0 and 1 such that $y_i$ takes the minimum of $1$ and $\sum\nolimits_{m \in {\cal B}}\sum\nolimits_{n \in {\cal B}^c}\sum\nolimits_{j \in \delta^+_i} x_{ij}^{mn}$.

While for constraint (\ref{op2-1b}), we first let
\begin{equation}\label{wn}
w_n = \min\limits_{m \in {\cal B}}\lambda_{m,n}({\mv X}) = \min\limits_{m \in {\cal B}} \biggl\{\sum\limits_{(i,j) \in {\tilde E}} c_{ij}x_{ij}^{mn}\biggl\}-1, \forall n \in {\cal B}^c,
\end{equation}
which transforms constraint (\ref{op2-1b}) into a linear constraint, i.e.,
\begin{equation}\label{eq4}
	\frac{1}{N}\sum\limits_{n \in \cal N}{w_n} \le \lambda_0.
\end{equation}
To linearize (\ref{wn}), we apply a similar ``big-$M$'' method as for constraint (\ref{op2-1a}) by introducing $N_{BS}$ auxiliary binary variables for each $w_n$ in (\ref{wn}), denoted as $z_{m,n}, m \in {\cal B}, n \in {\cal B}^c$. Then, (\ref{wn}) can be replaced by the following linear constraints without loss of optimality to (P2.1),
\begin{align}
&z_{m,n} \in \{0,1\}, \forall m \in {\cal B}, n \in {\cal B}^c, \label{eq5}\\
&\sum\limits_{(i,j) \in {\tilde E}} c_{ij}x_{ij}^{mn} - 1 - M(1-z_{m,n})\le w_n, \forall m \in {\cal B}, n \in {\cal B}^c, \label{eq6}\\
&\sum\limits_{(i,j) \in {\tilde E}} c_{ij}x_{ij}^{mn} - 1 \ge w_n, \forall m \in {\cal B}, n \in {\cal B}^c, \label{eq7}\\
&\sum\limits_{m \in \cal B}{z_{m,n}} = 1, \forall n \in {\cal B}^c. \label{eq8}
\end{align}
It is noted that (\ref{eq7}) restricts that each $w_n$ must be no larger than $\sum\nolimits_{(i,j) \in {\tilde E}} c_{ij}x_{ij}^{mn} - 1$ over $m \in {\cal B}$. However, due to (\ref{eq6}) and (\ref{eq8}), each $w_n$ must be no smaller than $\sum\nolimits_{(i,j) \in {\tilde E}} c_{ij}x_{ij}^{mn} - 1$ for some $m, m \in {\cal B}$. It thus follows that constraints (\ref{eq6}) and (\ref{eq7}) can only be satisfied at the same time when $w_n$ takes the minimum of $\sum\nolimits_{(i,j) \in {\tilde E}} c_{ij}x_{ij}^{mn} - 1$ over $m \in {\cal B}$.

Based on the above linearization procedures, (P2.1) can now be reformulated as the following MILP, i.e.,
\begin{align}
{\text{(P2.2)}}\; &\mathop {\min}\limits_{{\mv X}, {\mv Y},{\mv \rho},{\mv W},{\mv Z}}\;\; \sum\limits_{i \in {\cal B}^c} y_i \nonumber\\
\text{s.t.}\;\;& {\text{(\ref{flow}), (\ref{eq1})-(\ref{eq3})}}, {\text{(\ref{eq4})-(\ref{eq8})}}, \nonumber\\
&x_{ij}^{mn} \in \{0,1\}, \forall (i,j) \in {\tilde E}, m \in {\cal B}, n \in {\cal B}^c,
\end{align}
where ${\mv \rho}$, ${\mv W}$, and ${\mv Z}$ denote the ensemble of $\rho_i$'s, $w_n$'s and $z_{m,n}$'s, respectively.

Note that problem (P2.2) is an MILP which contains $(N-N_{BS})(\lvert \tilde E \rvert N_{BS}+N_{BS}+2)$ binary variables (in ${\mv X}$, ${\mv Y}$, ${\mv \rho}$, and ${\mv Z}$) and $N-N_{BS}$ non-binary variables (in $\mv W$). Thus, this problem can be optimally solved by applying the BB algorithm, which involves solving a sequence of linear programming problems. However, it is difficult to analyze the complexity of the BB algorithm for solving (P2.2), while the complexity should increase with the total number of cells (i.e., $N$) and edges in the LoS graph (i.e., $\lvert \tilde E \rvert$), as the size of the feasible set of (P2.2) increases with them. The worst-case complexity of the BB algorithm should amount to that of full enumeration, i.e., determining whether or not to deploy an IRS in each of the $N-N_{BS}$ cells in ${\cal B}^c$, which incurs computational complexity in the order of $2^{N-N_{BS}}$. However, the running time of the former is generally much less than that of the latter by properly discarding some solution sets that cannot yield the optimal solution to (P2.2).

\subsection{Low-Complexity Solution by Successive IRS Removal}
Although the optimal solution to (P2) can be obtained by the BB algorithm, its worst-case complexity, albeit rarely encountered, can still be high with increasing $N$ and/or $N_{BS}$. To address this issue, in this subsection, we propose a more efficient successive IRS removal algorithm to solve (P2). Its basic idea is to first deploy IRSs in all cells in ${\cal B}^c$ and then successively remove the IRSs in some cells, until no more IRS can be removed subject to the given performance constraint.

Specifically, we initialize the IRS deployment as ${\cal I}={\cal B}^c$, i.e., IRSs have been deployed in all cells without BS. In each iteration, we select the IRS in one cell and remove it if the average IRS number constraint in (P2) still holds after this removal. For example, consider that the IRS in cell $i, i \in {\cal B}^c$ is selected. If it is removed, then all edges starting from vertex $i$ should also be removed in $G$ or equivalently, their weights are set to $\infty$. Denote by ${\cal I}'={\cal I} \backslash \{i\}$ the resulting new IRS deployment. Then, we should check whether the condition $\lambda({\cal B},{\cal I}') \le \lambda_0$ is satisfied. To this end, the Dijkstra algorithm can be invoked to compute $\lambda_n({\cal B},{\cal I}')$ for each vertex $n, n \in {\cal B}^c$. If the above condition is satisfied, then we can safely remove the IRS in cell $i$. Otherwise, we need to select the IRS in another cell and check the above condition again.

\begin{figure}[!t]
\centering
\includegraphics[width=3in]{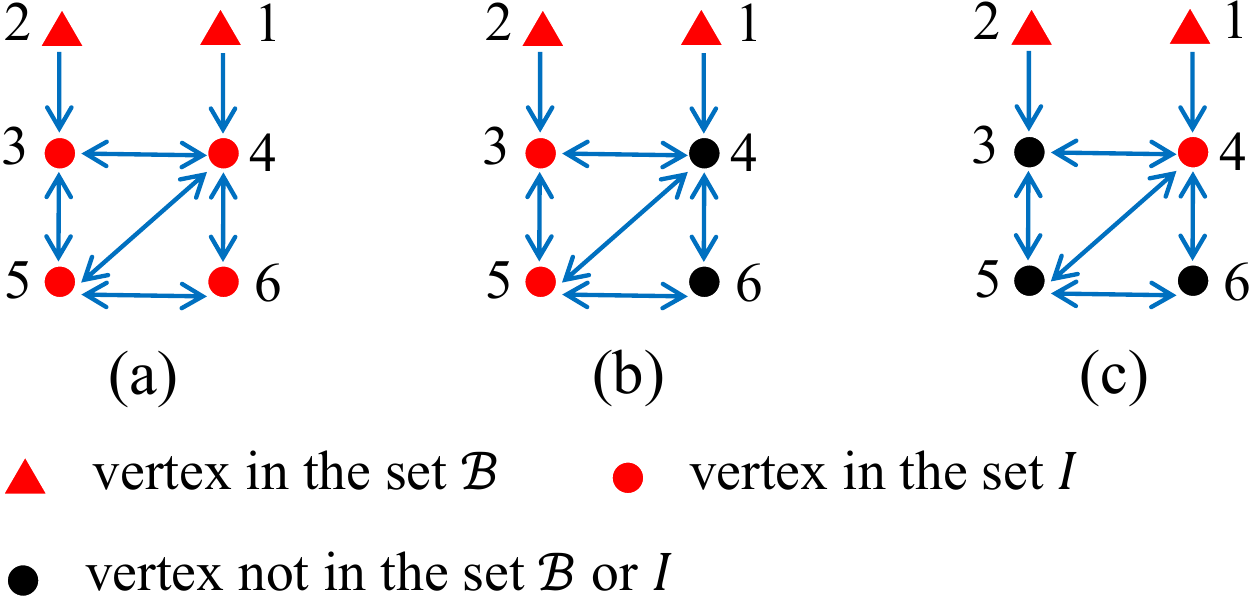}
\DeclareGraphicsExtensions.
\caption{Illustrative examples: (a) initial deployment with IRSs deployed in all cells (except cell 0) and converged deployments by successively removing the IRSs in (b) cells 4 and 6, and (c) cells 6, 5 and 3, respectively.}\label{example2}
\vspace{-6pt}
\end{figure}
By repeating the above process, the number of IRSs deployed, $\lvert \cal I \rvert$, can monotonically decrease with the iterations in the proposed algorithm, and the IRS removal can proceed until the convergence is reached. Nonetheless, it may converge to a suboptimal solution, and its ultimate performance depends critically on the order of selected cells for IRS removal. For example, for the graph $G$ in Fig.\,\ref{example2}(a) with ${\cal B}=\{1,2\}$ and ${\cal I}={\cal B}^c=\{3,4,5,6\}$, if we set $\lambda_0=2$ and successively remove the IRSs in cells 4 and 6, then the proposed algorithm would converge to ${\cal I}=\{3,5\}$, i.e., IRSs should be deployed in two cells, as shown in Fig.\,\ref{example2}(b). However, if we successively remove the IRSs in cells 6, 5 and 3, then IRSs only need to be deployed in one cell, i.e., ${\cal I}=\{4\}$, as shown in Fig.\,\ref{example2}(c). The reason is that cell 4 has a small minimum IRS number from the BS in cell 1, i.e., $\lambda_4({\cal B},{\cal B}^c)=0$ in Fig.\,\ref{example2}(a), and it can achieve LoS paths with three other cells 3, 5 and 6 or has three outgoing neighbors in $G$, i.e., $\lvert \delta^+_4 \rvert=3$. Thus, the IRS in cell 4 is involved in most of cascaded LoS paths in the region. However, in the former deployment case, the IRS in cell 4 is removed in the first iteration, thus significantly reducing the LoS path availability in the subsequent iterations.

Motivated by the above, under the IRS deployment $\cal I$ in each iteration of the proposed algorithm, we propose to select the IRSs/cells based on the descending order of their IRS numbers from their respectively associated BSs, i.e., $\lambda_{n}({\cal B},{\cal I}), n \in {\cal B}^c$. If more than one cells have the same value of $\lambda_{n}({\cal B},{\cal I})$, we further sort them based on the ascending order of their numbers of outgoing neighbors in $G$, i.e., $\lvert \delta^+_n \rvert = \sum\nolimits_{i \ne n}\mu_{n,i}, n \in \cal N$, so as to minimize the effect of IRS removal on the LoS path availability in $\cal D$. Based on the above criteria, the order of selection for all cells in $\cal I$ can be determined. For example, the first selected IRS is in cell $n_1$, with $n_1 = \arg\max\nolimits_{n \in \cal N}\lambda_{n}({\cal B},{\cal I})$. If $n_1$ is not unique, we should set $n_1 := \arg\min\nolimits_{n \in n_1}\lvert \delta^+_n \rvert$. The main procedures of the proposed algorithm are summarized in Algorithm 1. Note that the complexity of Algorithm 1 is mainly due to the use of the Dijkstra algorithm. For the LoS graph $G=(V,E)$, we have $\lvert V \rvert=N$ and $\lvert E \rvert=\sum\nolimits_{i=0}^N\sum\nolimits_{j \ne i} \mu_{i,j}$, and the Dijkstra algorithm incurs the complexity in the order of $\lvert V \rvert+\lvert E \rvert \log\lvert V \rvert$ per use, which is thus a polynomial function of $N$. As a result, the proposed Algorithm 1 is ensured to incur much lower complexity than the full enumeration which requires an exponential complexity over $N$. It will also be shown in Section \ref{sim} via simulation that Algorithm 1 can achieve a comparable performance to the BB algorithm but with much less running time.
\begin{algorithm}
  \caption{Successive IRS Removal Algorithm for Solving (P2)}\label{Alg1}
  \begin{algorithmic}[1]
    \State Initialize ${\cal I}={\cal B}^c$.
    \While {convergence is not reached}
    \State Invoke the Dijkstra algorithm to compute $\lambda_n({\cal I}), n \in {\cal I}$.
    \State Sort all vertices in ${\cal I}$ based on their $\lambda_n({\cal I})$'s and $\lvert \delta^+_n \rvert$'s, denoted as $\{n_1,n_2,\cdots,n_{\lvert \cal I \rvert}\}$.
    \State Let $i=1$.
    \While {$i \le \lvert \cal I \rvert$}
    \State Let ${\cal I}'={\cal I} \backslash \{n_i\}$ and compute $\lambda_n({\cal I}'), n \in {\cal N}$.
    \If {$\lambda({\cal I}') \le \lambda_0$}
    \State Update ${\cal I}={\cal I} \backslash \{n_i\}$.
    \State Go to line 3.
    \Else
    \State Update $i=i+1$.
    \EndIf
    \EndWhile
    \If {$i>\lvert \cal I \rvert$}
    \State Stop and output $\cal I$ as the deployment solution.
    \EndIf
    \EndWhile
  \end{algorithmic}
\end{algorithm}

\section{Proposed Solution to (P1) with Unknown BS Locations}\label{sol2}
In this section, we aim to solve the general problem (P1) with the BS locations unknown and to be optimized jointly with the IRS deployment, based on the proposed solution to (P2) in Section \ref{sol1} with given BS locations. Note that the optimal solution to (P1) can be obtained by enumerating all possible BS deployments; while for each BS deployment, the optimal BB algorithm in Section \ref{sol1} can be applied. Then, the optimal BS deployment (as well as the corresponding IRS deployment) can be obtained as the one with the minimum $\lvert \cal I \rvert$. However, this enumeration incurs the complexity in the order of $N \choose N_{BS}$, and the complexity of the BB algorithm also increases with $N$ and $N_{BS}$ in general. Thus, the complexity of obtaining the optimal solution to (P1) becomes considerably high if $N$ and/or $N_B$ is large.

To avoid the high enumeration complexity, we propose a new sequential update algorithm to solve (P1) iteratively, by sequentially updating the location of each BS while fixing those of all other BSs until no IRS number reduction can be achieved. Specifically, let ${\cal B}'=\{s_1,s_2,\cdots,s_{N_{BS}}\}$ denote the set of $N_B$ cells deployed with BSs. Consider that the deployment of the BS currently in cell $s_m$ needs to be updated now. Then, the following optimization problem should be solved,
\begin{align}
\mathop {\min}\limits_{s,{\cal I}}\;\;& \lvert {\cal I} \rvert \nonumber\\
\text{s.t.}\;\;&s \in  {\cal N} \backslash {\cal B}'_m, \nonumber\\
&{\cal I} \subseteq {\cal N} \backslash ({\cal B}'_m \cup s), \nonumber\\
&\lambda({\cal B}'_m \cup s,{\cal I}) \le \lambda_0.\label{sequential}
\end{align}
where ${\cal B}'_m \triangleq {\cal B}' \backslash \{s_m\}$ denotes the set of all cells except cell $s_m$.

For any given $s$, problem (\ref{sequential}) can be solved by applying one of the proposed algorithms in Section \ref{sol1}. In this paper, we apply the successive IRS removal algorithm as it has lower complexity than the BB algorithm. As such, we can enumerate all possible $N-N_{BS}$ cells in ${\cal N} \backslash {\cal B}'_m$ and select the one yielding the smallest objective value of problem (\ref{sequential}) as the optimized solution. Specifically, denote by $I_n$ the objective value of problem (\ref{sequential}) when fixing $s$ as the $n$-th cell in ${\cal N} \backslash {\cal B}'_m$. By letting $n^*=\arg \mathop {\min}\limits_{n} I_n$, the optimized solution of $s$ to (\ref{sequential}), denoted as $s^*$, should be determined as the $n^*$-th cell in ${\cal N} \backslash {\cal B}'_m$. Next, we can update ${\cal B}'$ by replacing $s_m$ therein with $s^*$, and the update for the next BS follows. It is not difficult to see that this process produces a non-increasing objective value of problem (\ref{sequential}) and thus, the convergence is guaranteed for the proposed sequential update algorithm. It is also worth mentioning that in the case of $N_{BS}=1$, the proposed algorithm is equivalent to a full enumeration of all possible BS deployment. Hence, it is able to obtain the optimal BS deployment solution when $N_{BS}=1$. Let $K$ denote the average number of updates per BS required for the proposed sequential update algorithm to converge (which is no larger than 2 based on our simulation setup in Section \ref{sim}). Then, its total number of enumerations is given by $KN_{BS}(N-N_{BS})$, which is much smaller than that required for finding the optimal BS deployment, i.e., $N \choose N_{BS}$, especially if $N$ and/or $N_{BS}$ is practically large.

However, the ultimate performance and total enumeration number of the proposed sequential update algorithm critically depend on the initial BS deployment. Particularly, if it is not properly initialized, the proposed algorithm may get trapped at undesired suboptimal deployment solutions or take a large enumeration number to converge. To obtain a high-quality initial BS deployment, we propose to first optimize the BS deployment to directly cover the maximum number of cells in $\cal D$ before the IRS deployment, which helps reduce the number of IRSs subsequently deployed to meet any performance target of $\lambda({\cal B},{\cal I})$. In particular, a cell $n, n \in \cal N$ can be directly covered if there exists at least one BS deployed in its incoming neighbor cells (i.e., $\delta^-_{n}$) and itself. Next, we show that the optimal BS deployment solution for coverage maximization can be obtained by formulating and solving an MILP.

Define $a_n \in \{0,1\}, n \in {\cal N}$ as a binary indicator which is equal to one iff cell $n$ is deployed with a BS. Furthermore, we define binary variables $b_n \in \{0,1\}, n \in {\cal N}$, which indicate that cell $n$ can be directly covered by at least one BS if $b_n=1$; otherwise, $b_n=0$. As a result, it can be easily seen that
\begin{equation}\label{bn}
b_n = \min \biggl\{1,\sum\limits_{m \in {\delta^-_{n} \cup \{n\}}}a_m\biggl\}, n \in {\cal B}.
\end{equation}
As a result, the total number of covered cells is given by $\sum\nolimits_{n \in {\cal N}}{b_n}$.

Let $\mv A$ and $\mv B$ denote the ensembles of $a_n$'s and $b_n$'s, respectively. Hence, the BS deployment problem for coverage maximization can be formulated as
\begin{align}
{\text{(P4)}}\; \mathop {\max}\limits_{{\mv A},{\mv B}}\;\;&\sum\limits_{n \in {\cal N}}{b_n} \nonumber\\
\text{s.t.}\;\;& \sum\limits_{n \in {\cal N}}{a_n}=N_{BS},\label{op4a}\\
& b_n \le \sum\limits_{m \in {\delta^-_{n} \cup \{n\}}}a_m, \forall n \in {\cal N},\label{op4b}\\
& a_n, b_n \in \{0,1\}, \forall n \in {\cal N},\label{op4c}
\end{align}
where constraint (\ref{op4a}) restricts that there are $N_{BS}$ BSs deployed in total, and we have replaced the equality constraint in (\ref{bn}) by the inequality constraint in (\ref{op4b}) without loss of optimality to (P4). The reason is that if $\sum\nolimits_{m \in {\delta^-_{n} \cup \{n\}}}a_m \ge 1$, since $b_n \in \{0,1\}$, the constraint in (\ref{op4b}) would become inactive, and the optimal $b_n$ should take the value of 1 to maximize the objective function of (P4). As a result, the equality in (\ref{bn}) still holds. On the other hand, if $\sum\nolimits_{m \in {\delta^-_{n} \cup \{n\}}}a_m=0$, since $b_n \in \{0,1\}$, $b_n$ can only take the value of 0, which satisfies (\ref{bn}) as well.

It is noted that (P4) is an MILP with $2N$ binary variables, which can be optimally solved by applying the BB algorithm. The main procedures of the proposed sequential update algorithm are summarized in Algorithm 2. It will be shown in Section \ref{sim} that Algorithm 2 can achieve near-optimal performance as compared to the full enumeration. In particular, the proposed initial BS deployment may be optimal for (P1) even without the need of sequential update, especially when $\lambda_0$ is set to be small. It is also worth noting that both Algorithms 1 and 2 can be executed offline and thus their complexity should be practically tolerable.
\begin{algorithm}
  \caption{Sequential Update Algorithm for Solving (P1)}\label{Alg2}
  \begin{algorithmic}[1]
    \State Solve (P4) via the BB algorithm to initialize ${\cal B}'$.
    \While {convergence is not reached}
    \State Let $m=1$.
    \While {$m \le N_{BS}$}
    \State Let $n=1$.
    \While {$n \le N-N_{BS}$}
    \State Fix $s$ as the $n$-th cell in the set ${\cal N} \backslash {\cal B}'_m$.
    \State Solve (\ref{sequential}) via Algorithm 1 and obtain $I_n$.
    \State Update $n=n+1$.
    \EndWhile
    \State Obtain $s^*$ as the $n^*$-th cell in ${\cal N} \backslash {\cal B}'_m$ with $n^*=\arg \mathop {\min}\limits_{n} I_n$.
    \State Update ${\cal B}'$ by updating $s_m$ as $s^*$.
    \State Update $m=m+1$.
    \EndWhile
    \EndWhile
    \State Output ${\cal B}'$ and its corresponding IRS deployment as the optimized solutions.
  \end{algorithmic}
\end{algorithm}

\section{Numerical Results}\label{sim}
\begin{figure}[!t]
\centering
\includegraphics[width=2.8in]{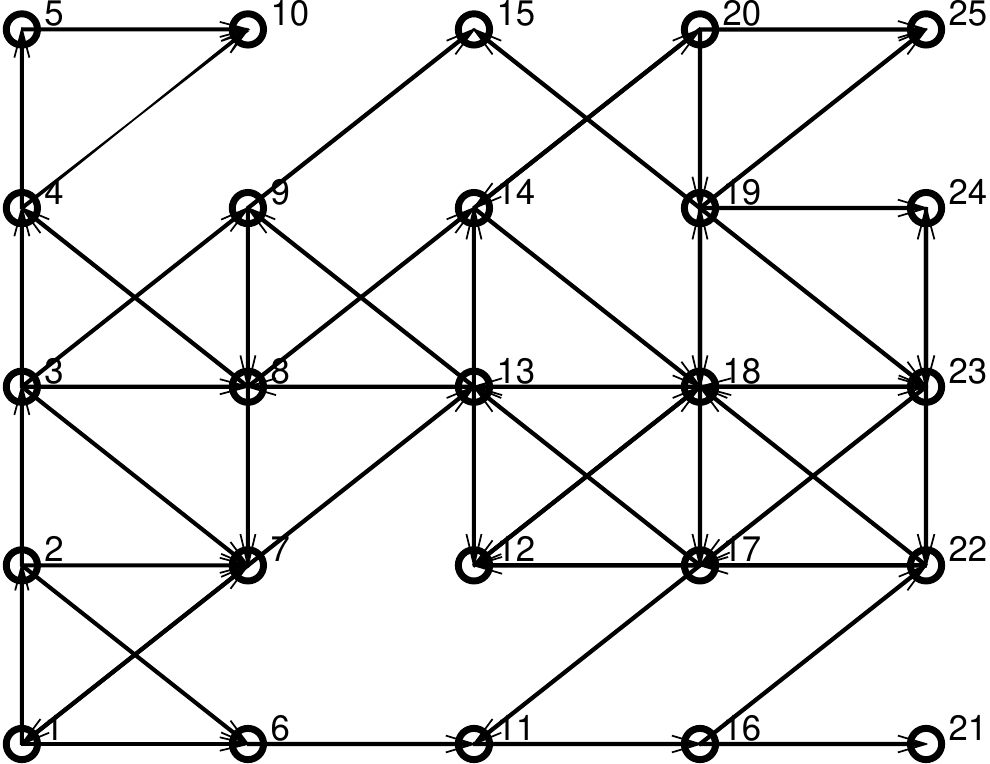}
\DeclareGraphicsExtensions.
\caption{LoS graph of the considered region.}\label{SimSetup}
\vspace{-6pt}
\end{figure}
In this section, numerical results are provided to demonstrate the effectiveness of the proposed algorithms for joint BS and IRS deployment. We consider a complex indoor environment which is divided into $N=25$ cells, and its corresponding LoS graph $G$ is shown in Fig.\,\ref{SimSetup}.

\subsection{Optimized IRS Deployment with Given BS Deployment}\label{sim1}
\begin{figure*}[hbtp]
\centering
\subfigure[$\lambda_0=1.88, N_{BS}=1$, 13 IRSs]{\includegraphics[width=0.4\textwidth]{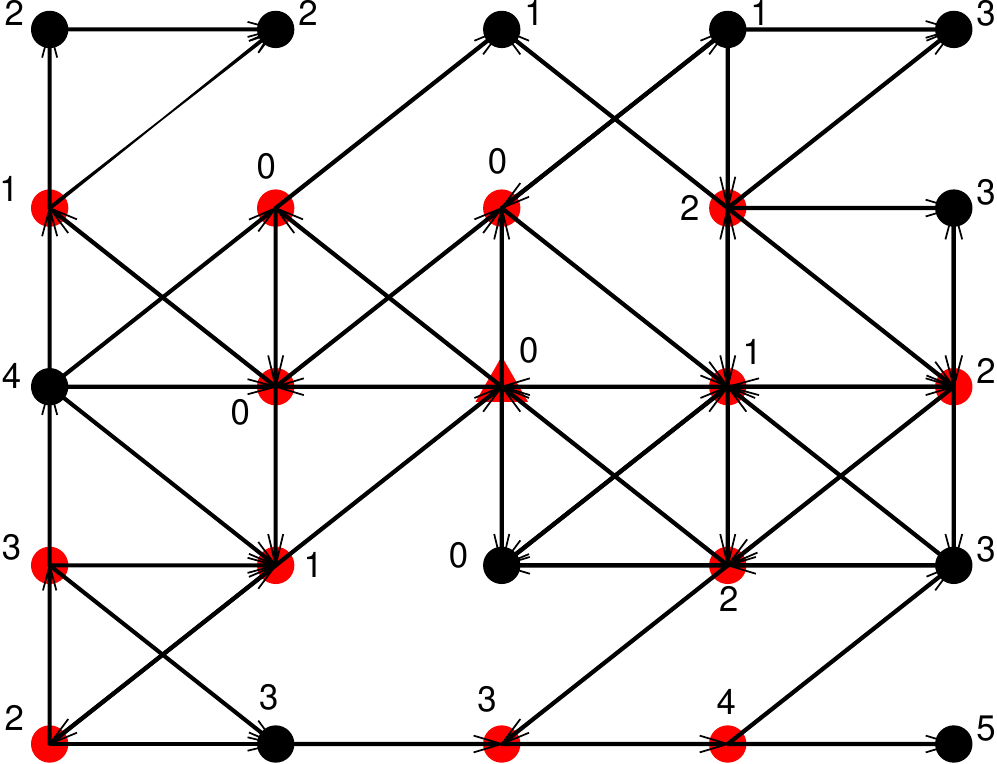}}\qquad
\subfigure[$\lambda_0=2.4, N_{BS}=1$, 11 IRSs]{\includegraphics[width=0.4\textwidth]{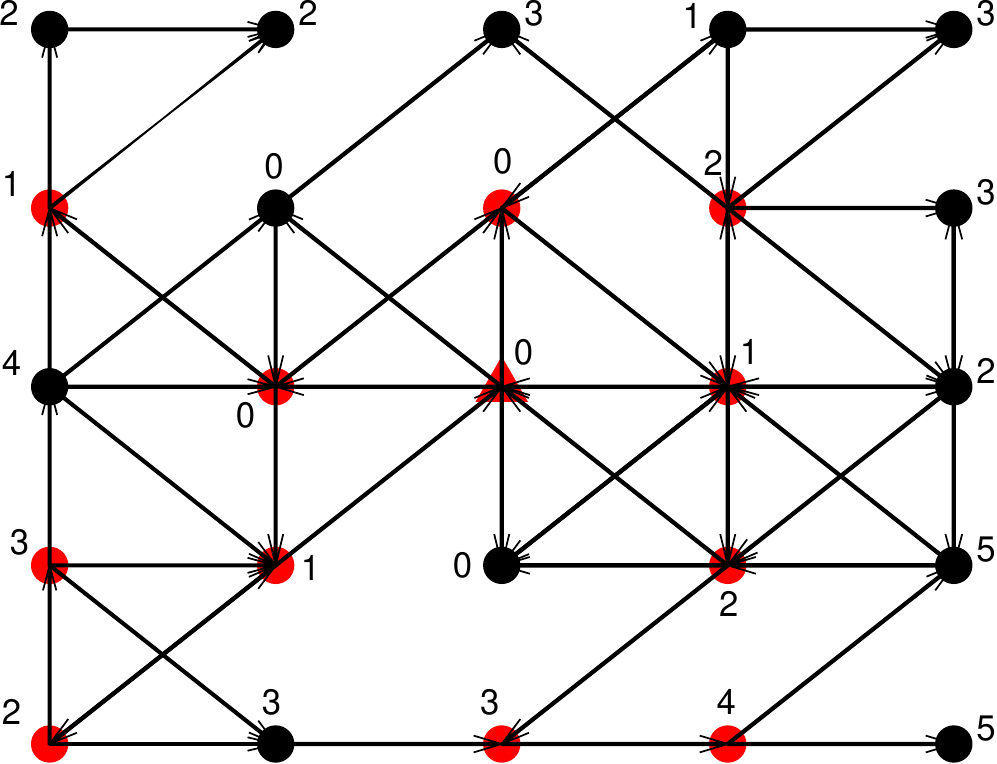}}\qquad
\subfigure[$\lambda_0=1.88, N_{BS}=2$, 9 IRSs]{\includegraphics[width=0.4\textwidth]{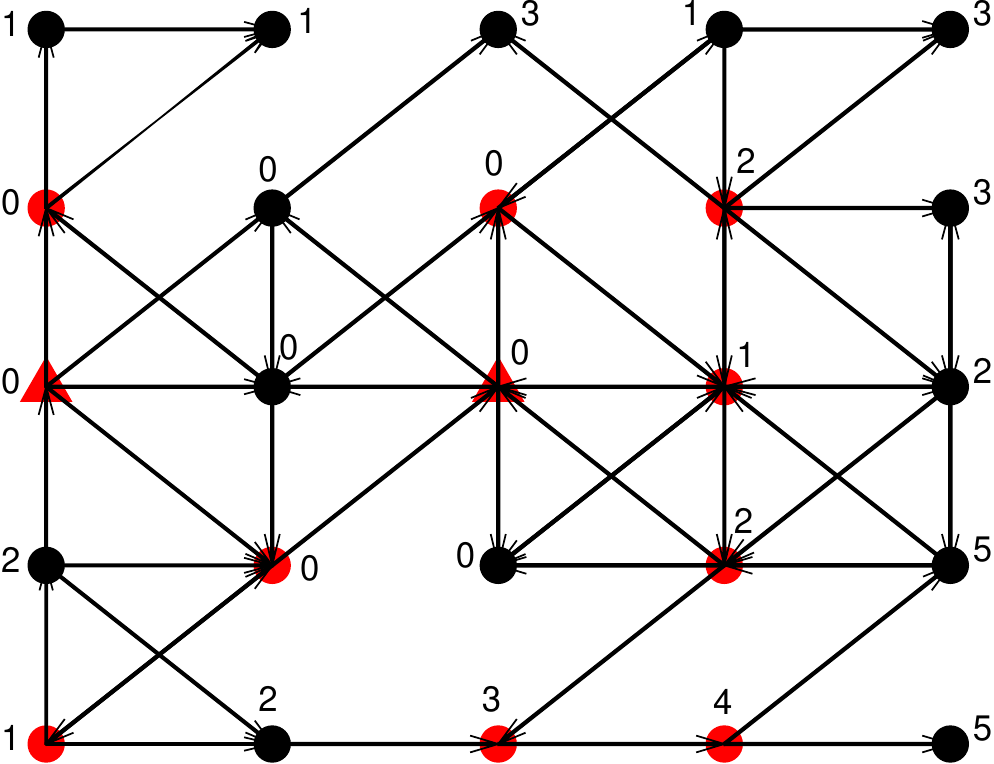}}\qquad
\subfigure[$\lambda_0=1.88, N_{BS}=3$, 8 IRSs]{\includegraphics[width=0.4\textwidth]{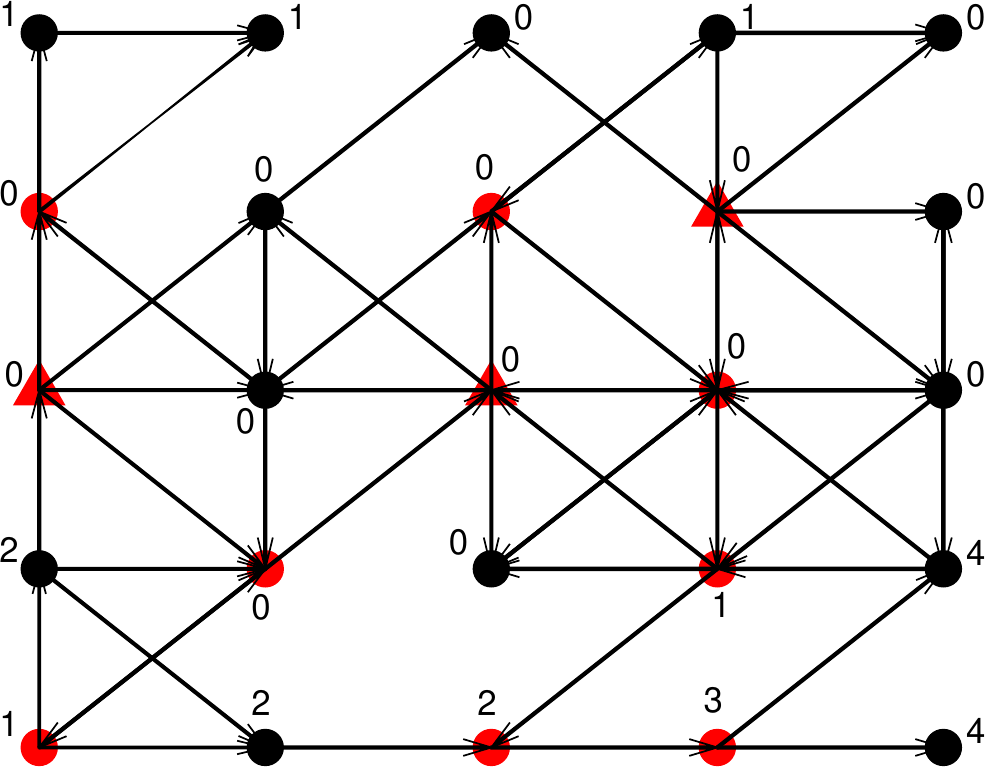}}\qquad
\caption{Optimal IRS deployment solutions to (P2) under different values of $\lambda_0$ and $N_{BS}$.}\label{IRSDeploy}
\vspace{-6pt}
\end{figure*}
First, Fig.\,\ref{IRSDeploy} plots the optimal IRS deployment solutions to (P2) by the BB algorithm under different values of $\lambda_0$ and $N_{BS}$. The locations of BSs are indicated by red triangles, while the cells with/without IRSs deployed are marked by red/black circles. The minimum IRS number for each cell $n, n \in \cal N$ from its associated BS, i.e., $\lambda_n({\cal B},{\cal I})$ in (\ref{min2}), is also shown. Figs.\,\ref{IRSDeploy}(a) and \ref{IRSDeploy}(b) show the optimal IRS deployment solutions under $\lambda_0=1.88$ and $\lambda_0=2.4$, respectively, with $N_{BS}=1$. The BS is assumed to be deployed in cell 13, i.e., ${\cal B}=\{13\}$. In this case, it can be computed that $\lambda({\cal B},{\cal B}^c)=1.88$. It is observed from Fig.\,\ref{IRSDeploy}(a) that the optimal solution to (P2) only deploys 13 IRSs in total to achieve the same $\lambda({\cal B},{\cal I})=\lambda({\cal B},{\cal B}^c)=1.88$ as deploying IRSs in all the 24 cells in ${\cal B}^c$. This implies that the proposed algorithm can help substantially reduce the IRS deployment cost in practice, without compromising the overall coverage performance. It is also observed from Fig.\,\ref{IRSDeploy}(b) that by increasing $\lambda_{0}$ from 1.88 to 2.4, the total number of deployed IRSs can be reduced from 13 to 11, which validates the performance-cost trade-off discussed in Section \ref{pm}. On the other hand, Figs.\,\ref{IRSDeploy}(c) and \ref{IRSDeploy}(d) show the optimal IRS deployment solutions to (P2) under $N_{BS}=2$ and $N_{BS}=3$, respectively, with $\lambda_0=1.88$. The BSs are assumed to be located in ${\cal B}=\{3,13\}$ and ${\cal B}=\{3,13,19\}$, with $\lambda({\cal B},{\cal B}^c)$ equal to 0.92 and 0.76, respectively, both of which are much smaller than that in the case of $N_{BS}=1$ with $\lambda({\cal B},{\cal B}^c)=1.88$. This implies that increasing the number of BSs deployed can significantly reduce the average minimum IRS number required per link. It is also observed from Figs.\,\ref{IRSDeploy}(c) and \ref{IRSDeploy}(d) that by increasing the number of BSs, the number of IRSs deployed can be dramatically reduced. As such, for any given $\lambda_0$, there also exists an inherent trade-off in balancing the number (or deployment cost) of BSs and IRSs deployed to minimize the total cost $c({\cal B},{\cal I})$ in (\ref{cost}), as will be shown later in Section \ref{sim3}.

Next, by varying the value of $\lambda_0$ in (P2), Fig.\,\ref{IRSNumVsHop} plots the achieved average minimum IRS numbers $\lambda({\cal B},{\cal I})$ versus the number of IRS deployed $\lvert {\cal I}\rvert$, under the BS deployment shown in Figs.\,\ref{IRSDeploy}(b) and \ref{IRSDeploy}(c), respectively. It is first observed that the proposed successive IRS removal algorithm can achieve the same performance as the optimal BB algorithm. It is also observed that there exist three boundary points in the optimal performance-cost trade-off region for both $N_{BS}=1$ and $2$. As expected, the average minimum IRS number is monotonically non-increasing with the number of IRSs deployed. However, as $\lvert{\cal I}\rvert \le 9$ and 11, the global LoS coverage with $\lambda({\cal B},{\cal I})<\infty$ cannot be achieved in the cases of $N_{BS}=2$ and $=1$, respectively. As $\lvert{\cal I}\rvert \ge 11$ and 13, the average minimum IRS number/coverage communication performance cannot be further decreased/improved by increasing the IRS deployment cost or number of IRSs in the considered region, when $N_{BS}=2$ and $=1$, respectively. Moreover, by increasing $N_{BS}$, the optimal trade-off region is observed to shrink, as expected. To further demonstrate the computational efficiency of the proposed successive IRS removal algorithm, we compare its running time (in second) with that of the BB algorithm in solving (P2) in Table \ref{time}, with $N_{BS}=1, 2$ and $3$. It is observed that the successive removal algorithm takes much less running time than the BB algorithm to obtain a near-optimal solution to (P2), and the saving in time becomes more significant as $N_{BS}$ increases.\vspace{-12pt}
\begin{figure}[!t]
\centering
\includegraphics[width=3.4in]{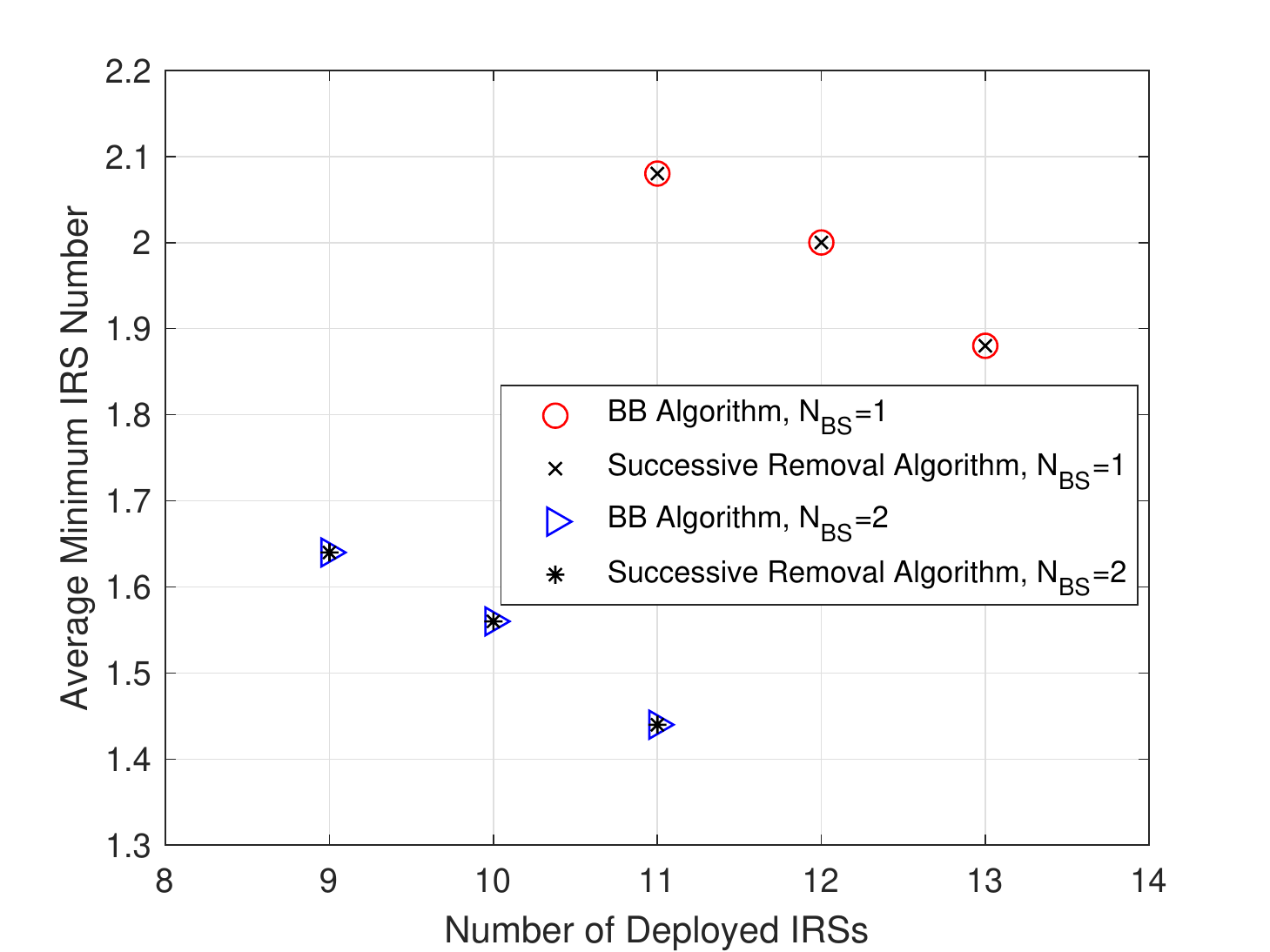}
\DeclareGraphicsExtensions.
\caption{Average minimum IRS number versus total deployment cost with given BS deployment.}\label{IRSNumVsHop}
\vspace{-6pt}
\end{figure}
\begin{table}[!t]
\centering
\small\caption{Running time in second of different algorithms for solving (P2)}\label{time}
\begin{tabular}{|c|c|c|c|}\hline
                  & $N_{BS}=1$ & $N_{BS}=2$ & $N_{BS}=3$  \\ \hline
BB algorithm      &  0.22   &   1.52   &   8.53 \\ \hline
\makecell{Successive IRS\\ removal algorithm}      &  0.11   &   0.23   &   0.32 \\ \hline
\end{tabular}
\vspace{-9pt}
\end{table}

\subsection{Jointly Optimized BS and IRS Deployments}
\begin{figure}[hbtp]
\centering
\subfigure[$N_{BS}=1$, 11 IRSs]{\includegraphics[width=0.4\textwidth]{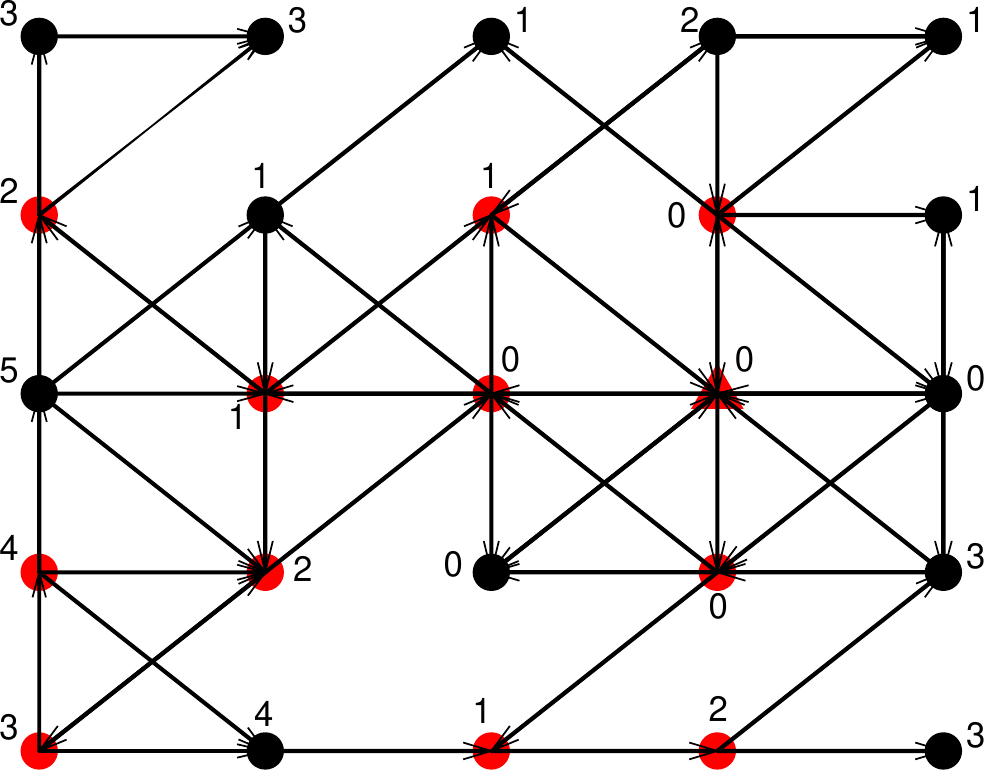}}\qquad
\subfigure[$N_{BS}=2$, 8 IRSs]{\includegraphics[width=0.4\textwidth]{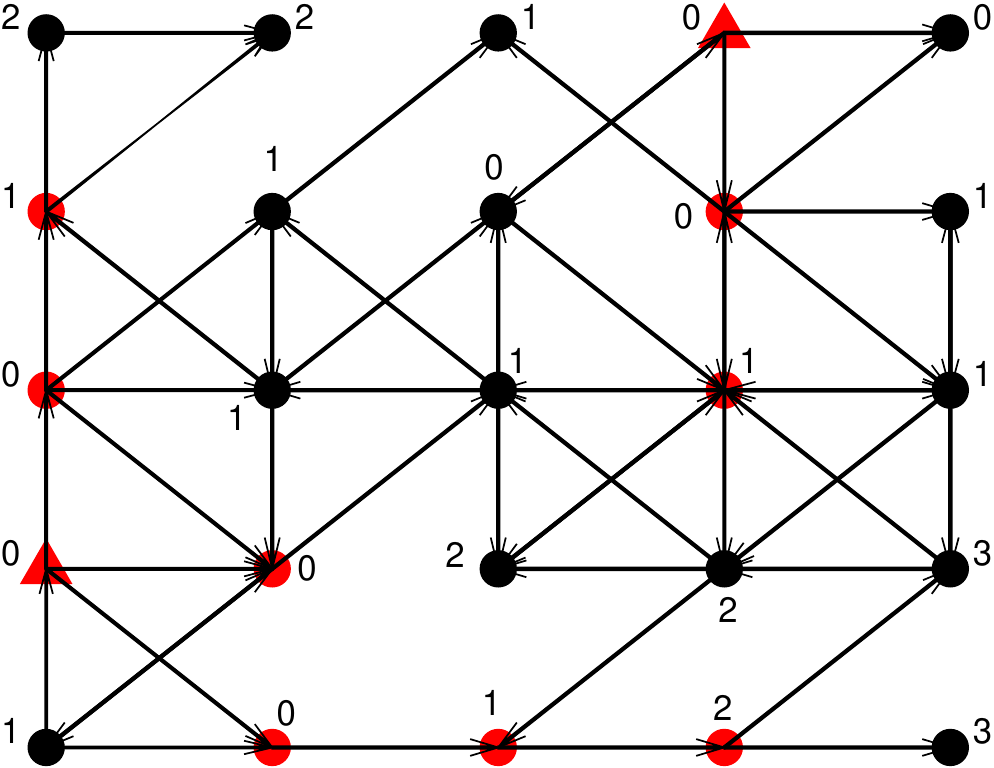}}\qquad
\caption{Optimized BS and IRS deployment solutions to (P1) under $\lambda_0=1.88$.}\label{JODeploy}
\vspace{-9pt}
\end{figure}
In Fig.\,\ref{JODeploy}, we plot the optimized BS and IRS deployment solutions to (P1) by the proposed subsequent update algorithm under $\lambda_0=1.88$. By comparing Fig.\,\ref{JODeploy} with Fig.\,\ref{IRSDeploy}, it is observed that by jointly optimizing the BS and IRS deployment, the number of IRSs deployed or IRS deployment cost can be further decreased, without the need of increasing the number of BSs. In particular, when $N_{BS}=1$ and $N_{BS}=2$, the numbers of IRSs deployed can be reduced from 13 and 9 in Fig.\,\ref{IRSDeploy} to 11 and 8 in Fig.\,\ref{JODeploy}, respectively. It is also observed that as compared to Fig.\,\ref{IRSDeploy}(c), the optimized locations of the two BSs are more separated, which helps cover a larger portion of the region with fewer IRSs.

\begin{figure}[!t]
\centering
\includegraphics[width=3.4in]{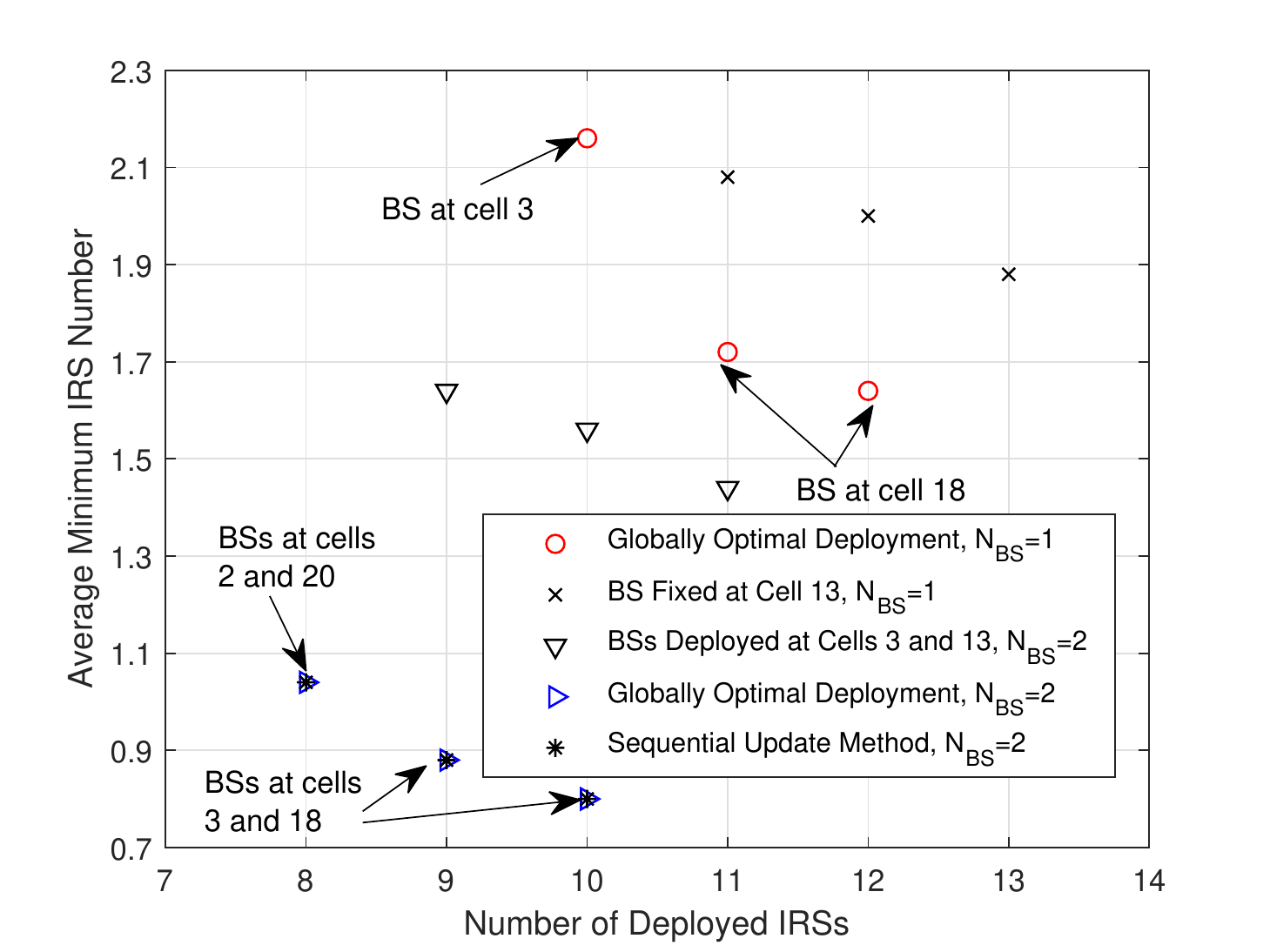}
\DeclareGraphicsExtensions.
\caption{Average minimum IRS number versus total deployment cost with optimized BS deployment.}\label{IRSNumVsHopJO}
\vspace{-9pt}
\end{figure}
Next, Fig.\,\ref{IRSNumVsHopJO} plots the achieved average minimum IRS numbers $\lambda({\cal B},{\cal I})$ versus the number of IRS deployed $\lvert {\cal I}\rvert$ with the optimized BS deployment. The following main observations can be made from Fig.\,\ref{IRSNumVsHopJO}. It is first observed that in the joint BS and IRS deployment optimization case, the optimized BS deployment may not belong to the same set of cells as the number of IRSs deployed changes. For example, when $N_{BS}=1$, the optimal BS deployment is at cells 3 and 18 when the number of IRSs is 10 and 11, respectively. Second, it is observed that as compared to the fixed BS deployment, the joint BS and IRS deployment helps improve the performance-cost trade-off by exploiting more design degrees of freedom. In particular, when $N_{BS}=1$ and $N_{BS}=2$, the former deployment requires 11 and 9 IRSs to achieve the global LoS coverage, while the latter deployment only requires 10 and 8 IRSs to achieve this purpose, respectively. Third, it is observed that the proposed sequential update algorithm can achieve the same performance as the optimal BS and IRS deployment solution obtained from the full enumeration. Particularly, the initial BS deployment solutions by the proposed algorithm in Section \ref{sol2} are given by ${\cal B}=\{18\}$ and ${\cal B}=\{3,18\}$ in the cases of $N_{BS}=1$ and $N_{BS}=2$, respectively, both of which are observed to be indeed optimal when the desired average minimum IRS number is small.

\subsection{Trade-off in the Numbers of BSs and IRSs}\label{sim3}
\begin{figure}[!t]
\centering
\includegraphics[width=3.4in]{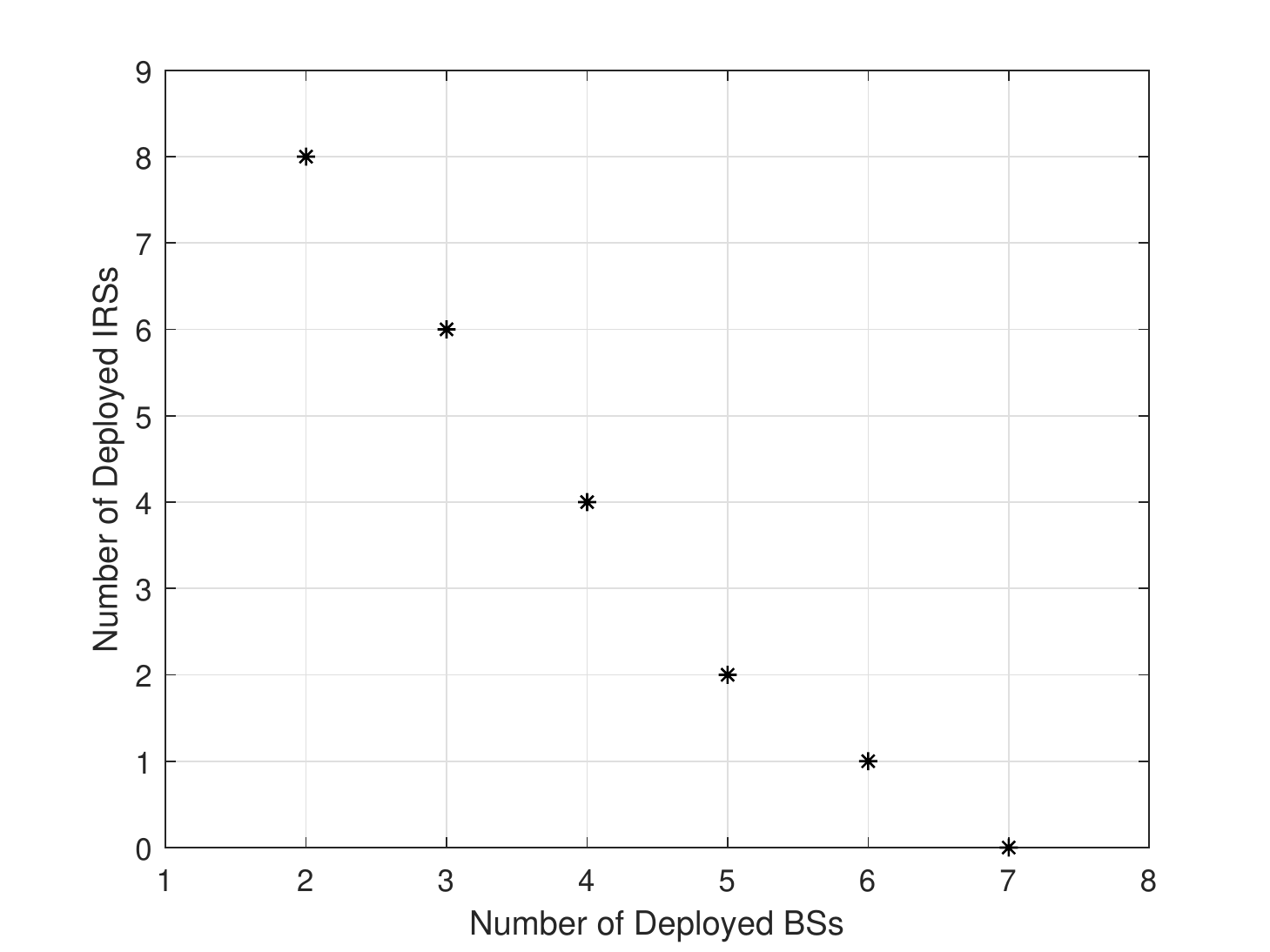}
\DeclareGraphicsExtensions.
\caption{Optimized BS and IRS number region with $\lambda_0=1.2$.}\label{NumRegion}
\vspace{-9pt}
\end{figure}
As previously discussed in Section \ref{sim1}, for any given $\lambda_0$, there exists an optimal pair of BS and IRS numbers to minimize the total cost $c({\cal B},{\cal I})$ in (\ref{cost}), depending on the ratio of $\alpha_B$ to $\alpha_I$. To show this trade-off, we plot in Fig.\,\ref{NumRegion} the optimized number of IRSs, $\lvert {\cal I} \rvert$, in solving (P1) versus the number of BSs, $N_{BS}$, under $\lambda_0=1.2$. As expected, it is observed that increasing $N_{BS}$ can help reduce the number of IRSs deployed, as this results in a larger size of $\cal B$ in (\ref{min2}). In particular, when $N_{BS} \ge 7$, there is no need to deploy any IRS to achieve the global LoS coverage over $N=25$ cells. Given the six boundary points in the trade-off region shown in Fig.\,\ref{NumRegion}, it can be calculated that if $1 < \alpha_B/\alpha_I < 2$, the optimal BS and IRS number pair that minimizes $c({\cal B},{\cal I})$ is $(5,2)$. While if $\alpha_B/\alpha_I \ge 2$, it is given by $(2,8)$. This implies that if the deployment cost per BS is increasingly higher than that per IRS, BSs should be minimally deployed in order to minimize the total deployment cost $c({\cal B},{\cal I})$.

\section{Conclusions}
In this paper, we study a new joint BS and IRS deployment problem for enhancing the wireless network coverage performance by exploiting the multi-IRS LoS reflections. For a given region, we first propose a graph-based system model for multi-BS and multi-IRS deployment and characterize a fundamental performance-cost trade-off in this design problem. An optimal algorithm and two suboptimal algorithms of lower complexity are then proposed to solve the design problem efficiently with or without known BS locations, by dispensing with a full enumeration of all BS/IRS deployment solutions. Numerical results validate the performance-cost trade-off and show that this trade-off can be improved by optimizing the BS deployment in addition to the IRS deployment. Both proposed suboptimal algorithms are shown able to achieve near-optimal performance for multi-BS/IRS deployment in practice. It is also revealed that the number of BSs and IRSs needs to be optimally chosen to minimize the total deployment cost for any given communication requirement. This paper can be extended in various promising directions for future work. For example, we only focus on LoS coverage in this paper and consider the average minimum IRS number per link as the communication performance metric. In fact, the coverage and communication performance can be more accurately characterized by taking into account e.g., the statistics of the BS-IRS and inter-IRS channels, which can be leveraged to further improve the BS/IRS deployment design.

\bibliography{IRSdeploy}
\bibliographystyle{IEEEtran}

\end{document}